\documentclass[epj,nopacs]{svjour}

\RequirePackage[switch*]{lineno}

\usepackage{graphicx}
\usepackage{amssymb}

\usepackage{atlasphysics}
\usepackage{subfigure}
\usepackage{longtable}
\usepackage{multirow}
\usepackage{authblk}
\usepackage{booktabs}
\usepackage{cite}
\usepackage{amsmath}
\usepackage{bm}
\usepackage{verbatim}
\usepackage{hyperref}

\newcommand{\beq}{\begin{equation}}
\newcommand{\eeq}{\end{equation}}
\newcommand{\bea}{\begin{eqnarray}}
\newcommand{\eea}{\end{eqnarray}}
\newcommand{\tref}[1]{Table~\ref{#1}}
\newcommand{\sref}[1]{Section~\ref{#1}}
\newcommand{\cref}[1]{Chapter~\ref{#1}}

\def\mT{\ensuremath{m_{\mathrm{T}}}} 
\newcommand{\ggF}{\ensuremath{gg\to H}}
\newcommand{\VBF}{\ensuremath{qq\to qqH}}
\newcommand{\VH}{\ensuremath{q\bar{q}\to WH/ZH}}
\newcommand{\ttH}{\ensuremath{q\bar{q}/gg \to t\bar{t}H}}
\newcommand{\Zgjets}{$Z/\gamma^{*}$+jets}
\newcommand{\Zjets}{$Z$+jets}
\newcommand{\mll}{$m_{\ell\ell}$}
\newcommand{\dphill}{$\Delta\phi_{\ell\ell}$}

\newcommand{\hgg}{\ensuremath{H \rightarrow \gamma\gamma}}
\newcommand{\hww}{\ensuremath{H \rightarrow WW}}
\newcommand{\hwwlnln}{\ensuremath{H \rightarrow WW^{(*)} \rightarrow \ell\nu\ell\nu}}
\newcommand{\hwwlnqq}{\ensuremath{H \rightarrow WW \rightarrow \ell\nu qq}}
\newcommand{\hzz}{\ensuremath{H \rightarrow ZZ}}

\newcommand{\hZZllnn}{\ensuremath{H \rightarrow ZZ\rightarrow \ell\ell\nu\nu}}
\newcommand{\hZZllqq}{\ensuremath{H \rightarrow ZZ\rightarrow \ell\ell qq}}
\newcommand{\hZZllll}{\ensuremath{H \rightarrow ZZ^{(*)}\rightarrow \ell\ell\ell\ell}}

\newcommand{\lumiuncertainty}{$\pm3.4$\%}
\newcommand{\lhood}{\ensuremath{{\cal L}}}

\newcommand{\ZZbkg}{\ensuremath{ZZ^{(*)}}}
\newcommand{\Wjets}{$W$+jets}
\newcommand{\SM}{Standard Model}

\begin{document}
\switchlinenumbers

\title{Limits on the production of the  Standard Model Higgs Boson in $pp$ collisions at $\sqrt{s}$~=~7~TeV with the ATLAS detector}
\author{The ATLAS collaboration
\\ CERN-PH-EP-2011-076\\Submitted to EPJC}
\institute{}
\date{July 16, 2011}
\abstract{
A search for the \SM\ Higgs boson  at the Large Hadron
  Collider (LHC) running at a centre-of-mass 
energy of 7~\TeV\ is reported, based on a total integrated luminosity of
up to 40~\ipb\ collected by the ATLAS detector in 2010. 
Several Higgs boson decay
channels: 
\hgg, \hZZllll, \hZZllnn, \hZZllqq, \hwwlnln\ and \hwwlnqq\ ($\ell$ is
e, $\mu$) are
combined in a mass range from 110~GeV to 600~GeV.  
The highest sensitivity is achieved in the mass range
between 160~GeV and 170~GeV, where the expected 95\% CL exclusion sensitivity 
is at Higgs boson production
cross sections
 2.3 times the Standard Model prediction.
Upper limits on the cross section for its production are determined.
Models with a fourth generation 
of heavy leptons and quarks with \SM-like couplings to the Higgs boson
are also investigated and are excluded at 95\% CL 
for a Higgs boson mass in the range from 140 GeV to 185 GeV.\\ 
}
\authorrunning{ATLAS collaboration}
\titlerunning{Search for the Standard Model Higgs Boson}

\maketitle


\section{Introduction}
\label{sec:intro}

The search for the \SM\ Higgs
boson~\cite{Englert:1964et,Higgs:1964ia,Guralnik:1964eu}
is one of the key aims of the Large Hadron Collider (LHC) at 
CERN. Prior to the LHC, the best direct information is  a lower limit
of 114.4~\GeV, set using the combined results of the four  LEP experiments~\cite{Barate:2003sz}, and an excluded band of
158~\GeV\ to 173~\GeV\ from the combined Tevatron
experiments~\cite{:2010ar,ref:tevhiggs}. 
First results from the ATLAS experiment are available in various \SM\
Higgs boson search
channels~\cite{hgg,hwwlnln,hwwlnqq,hzzllll,hzzllqq}. 
There are also results from
the CMS collaboration~\cite{Chatrchyan:2011tz} in the
\hwwlnln\footnote{In this paper, 
  the raised index '*' implies 
  a particle off mass-shell, $\ell$ is always taken to mean either $e$ or
  $\mu$ and $q$ can be any of $u$, $d$, $s$, $c$ or $b$. } channel which have
a sensitivity
similar to  the equivalent search reported here.
These results are based on proton-proton collision data
collected in 2010 at a 
centre-of-mass energy of $\sqrt{s}=7$~TeV. 
This paper combines the results from the different Higgs boson searches
to obtain the overall
sensitivity to a \SM\ Higgs boson with the 2010 ATLAS dataset. 

All analyses use the most detailed  calculations available for 
the cross sections, as discussed in \sref{sec:higgsxsbr}. 
The searches in individual Higgs boson decay channels $H\to
\gamma\gamma$, $H\to WW^{(*)}$
 and $H\to ZZ^{(*)}$ are outlined in 
Sections~\ref{sec:hgamgam},~\ref{sec:hww}, and~\ref{sec:hzz},
respectively.  The statistical interpretation
  uses the profile-likelihood
ratio~\cite{cscbook} as test-statistic. 
Thirty-one Higgs boson masses, in steps of 10~\GeV\ from 110~\GeV\ to
200~\GeV\ (plus 
115~\GeV\ in addition) and 20~\GeV\ from 200~\GeV\ to 600~\GeV, are tested.  
Exclusion limits are
obtained using the power constrained CL$_{\mathrm sb}$ limit~\cite{ref:pcl},  as
discussed in \sref{sec:stat}. To allow for comparisons with the exclusion limits
obtained by other experiments, the results are also determined using the CL$_s$
method~\cite{Read:2002hq}. The limits are presented in terms of 
$\sigma/\sigma_{\mathrm SM}$, the multiple 
of the expected \SM\ cross section at the Higgs boson mass considered. Results are
also presented in terms 
of the corresponding ratio where the cross section in the denominator
includes the effects of a fourth generation of heavy leptons and quarks with \SM-like
couplings to the Higgs boson. 
Section~\ref{sec:systematics} describes the treatment of the major sources of systematic
uncertainty in the combined likelihood. The limits for individual channels and the 
combined results are detailed in \sref{sec:combination} and the conclusions are drawn in \sref{sec:conc}.

\section{The ATLAS Detector}
\label{sec:detector}
The ATLAS experiment~\cite{Aad:2008zzm} is a multipurpose particle
physics apparatus with forward-backward symmetric cylindrical
geometry covering $|\eta|<2.5$ for tracks and $|\eta<4.5$ for jets\footnote{ATLAS uses a right-handed coordinate system with its
origin at the nominal interaction point (IP) in the centre of the detector and the $z$-axis coinciding with the axis of the beam pipe. The $x$-axis points from the IP to the centre of the LHC ring, and the $y$-axis points upward. Cylindrical coordinates $(r, \phi)$ are used in the transverse plane, $\phi$ being the azimuthal angle around the beam pipe. The pseudorapidity is defined in terms of the polar angle $\theta$ as $\eta = -\ln \tan(\theta/2)$.}. 
The inner tracking detector (ID) consists of a silicon pixel detector, a silicon microstrip detector (SCT), and a transition radiation tracker (TRT). The ID is surrounded by a thin superconducting solenoid providing a 2 T 
magnetic field, and by high-granularity liquid-argon (LAr) sampling
electromagnetic calorimeters. An iron-scintillator tile calorimeter
provides hadronic coverage in the central rapidity range. The end-cap
and forward regions are instrumented with LAr calorimetry for both
electromagnetic and hadronic measurements. The muon spectrometer (MS)
surrounds the calorimeters and consists of three large superconducting
toroids, each with eight coils, a system of precision tracking
chambers, and detectors for 
triggering. 

The data used in this analysis were recorded in 2010 at the LHC at a centre-of-mass energy of 7 TeV. 
 Application of beam, detector, and data-quality requirements results
 in a total integrated luminosity of 35 to 40~\ipb\ depending on the
 search channel, with an estimated uncertainty of
 \lumiuncertainty~\cite{atlas-update-lumi-conf}. The events were 
 triggered either by a single lepton or a pair of photon
 candidates with  transverse momentum (\pt)  thresholds which were
 significantly below the 
 analysis offline requirements. The trigger introduces very little
 inefficiency  except in one channel, \hwwlnqq, where there are
 efficiency losses   in the muon channel  of about  16\%.

Electron and photon candidates are reconstructed from energy clusters
recorded
in the liquid-argon  electromagnetic calorimeter.
The clusters must have shower profiles
consistent with those expected from an electromagnetic shower.
Electron candidates are  matched to tracks reconstructed in the inner detector, while
photon candidates  require either no track or an identified
conversion candidate.  Muon candidates are reconstructed by matching
tracks found in the inner detector with either tracks or hit segments
in the muon spectrometer.  Details of the quality criteria required on
each of these objects differ amongst the analyses discussed
here. There are in addition isolation criteria which again depend upon
the specific backgrounds relevant to each  analysis.

Jets are reconstructed from topological clusters\cite{:2010wv} in the
calorimeter using an anti-k$_{\mathrm t}$ algorithm\cite{Cacciari:2008gp} with a radius
parameter R = 0.4. They are calibrated
~\cite{:2010wv,ATL-CONF-2011-032}  
from the electromagnetic scale 
to the hadronic energy scale using  \pT\ and $\eta$ dependent
correction factors based on Monte Carlo simulation and validated on data. They
are required to have a \pT\ greater than 25~\GeV\ unless otherwise
stated.
B tagging is performed using a secondary vertex algorithm based upon
the decay length significance. A selection requirement is set to
describe a jet as 
`$b$-tagged' which has a 50\% efficiency for true $b$-jets. 
The missing transverse energy
 is reconstructed  from topological
energy clusters in the ATLAS calorimeters, with corrections for
measured muons.

\section{Cross sections, decays and simulation tools}
\label{sec:higgsxsbr}

\subsection{Search for the \SM\ Higgs boson}

At the LHC, the most important \SM\ Higgs boson production processes
are  the following four: gluon fusion (\ggF),  
which couples to the Higgs boson via a heavy-quark triangular loop;
fusion of vector bosons radiated off quarks (\VBF); 
associated production with a vector boson (\VH); 
associated production with a top-quark pair (\ttH). 
The current calculations of the production cross sections have been gathered
and summarised in Ref.~\cite{LHCHiggsCrossSectionWorkingGroup:2011ti}. 

Higher-order corrections have been calculated up to next-to-next-to-leading order (NNLO) in QCD 
for the gluon fusion~\cite{Harlander:2002wh,Anastasiou:2002yz,Ravindran:2003um,Anastasiou:2008tj,deFlorian:2009hc,Baglio:2010ae}, 
vector boson fusion~\cite{Bolzoni:2010xr}
and associated $WH$/$ZH$ production processes~\cite{Brein:2003wg}, 
and to next-to-leading order (NLO) for the associated 
production with a \ttbar\ 
pair~\cite{Beenakker:2001rj,Dawson:2002tg}. 
In addition, QCD soft-gluon resummations up to next-to-next-to-leading log (NNLL) are  available for the 
gluon fusion process~\cite{Catani:2003zt}. 
The NLO electroweak (EW) corrections are applied to the gluon fusion~\cite{Aglietti:2004nj,Actis:2008ug}, 
vector boson fusion~\cite{Ciccolini:2007jr,Ciccolini:2007ec} 
and the associated $WH$/$ZH$ production~\cite{Ciccolini:2003jy} processes.

The Higgs boson decay branching ratios used take  into
account the recently calculated higher order QCD and EW corrections in
each Higgs boson decay
mode~\cite{Djouadi:1997yw,LHCHiggsCrossSectionWorkingGroup:2011ti}. The
errors in these calculations for the states considered here are at
most 2\% and are neglected.
For most four-fermion final states
the predictions by Prophecy4f~\cite{Bredenstein:2006rh,Bredenstein:2006ha} 
are used which include the complete NLO QCD+EW
corrections with all interference and leading two-loop heavy Higgs
boson corrections to the four-fermion width.  
The $\hZZllqq$ and \hZZllnn\ analyses use the less precise single $Z$
boson decay rates 
from Ref.~\cite{Nakamura:2010zzi}. %

\begin{figure}[htp]
\begin{center}
\includegraphics[width=0.49\textwidth]{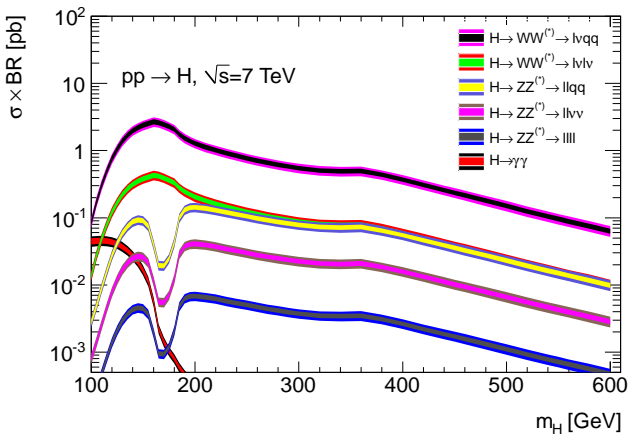}
\caption{The cross section multiplied by decay branching
  ratios  for Standard Model Higgs boson  
  production in $pp$ collisions at a 7 TeV centre-of-mass
  energy as a function
  of mass\cite{LHCHiggsCrossSectionWorkingGroup:2011ti}. 
All production modes are summed, and
only final states considered in this paper are shown.
Two bands are shown for each curve; the inner represents the QCD scale
uncertainty and the outer also includes the $\alpha_s$ and PDF uncertainty.
\label{fig:sm-crossbr} }
\end{center}
\end{figure}

The total signal production cross section 
in $pp$ collisions at \mbox{$\sqrt{s}=7$}~\TeV, multiplied by the
branching ratio for the final states considered in this paper,  is
summarised in Fig.~\ref{fig:sm-crossbr}  as a  function of the Higgs
boson mass.  
Sources of uncertainties on these cross sections include missing
higher-order corrections, imprecise knowledge of the parton
distribution functions (PDFs) and the uncertainty on the strong
force coupling constant, $\alpha_s$.
These uncertainties are treated according to the recommendations given in 
Refs.~\cite{LHCHiggsCrossSectionWorkingGroup:2011ti,Botje:2011sn,Ball:2011mu,Lai:2010vv,Martin:2009iq} and are $\pm(15$-$20)\%$ for the gluon
fusion process, $\pm(3$-$9)\%$ for the vector boson fusion process and $\pm5\%$ for the associated $WH$/$ZH$ production process.

\subsection{Higgs boson search in fourth generation models}

Models with a fourth generation of heavy leptons and quarks with
\SM-like couplings to the Higgs boson enhance its
production cross section in gluon fusion by a factor of 4 to 10
compared to the predicted 
rate with three generations~\cite{Kribs:2007nz,Schmidt:2009kk,Anastasiou:2010bt,Li:2010fu}. 
The model considered here~\cite{ruan_zhang}  has very heavy fourth
generation fermions, 
giving a minimum cross section but excluding the possibility that the
Higgs boson decays to 
heavy neutrinos.
These can weaken the exclusion for
Higgs boson masses below the W pair threshold\cite{Rozanov:2010xi}.
It should be noted that the branching ratio into photons is suppressed
by a factor around 8 in this model.

The Higgs boson production cross section in the gluon fusion process and
its decay branching ratios  
 have been calculated in the fourth generation model at NLO with HIGLU \cite{Spira:1995mt} and HDECAY~\cite{Djouadi:1997yw}.
The NNLO+NNLL QCD corrections are applied to the gluon fusion cross sections. 
The QCD corrections for the fourth generational model are assumed to be the
same as in the Standard Model.
The full two-loop \SM\ electroweak corrections~\cite{Aglietti:2004nj,Actis:2008ug} are taken into account.
The effect of a fourth generation in the \SM\ background processes,
which includes contributions from loop diagrams, has been neglected.

\subsection{Monte Carlo simulations}

For the H$\ra$ZZ Monte Carlo samples, the Higgs signal is generated
using PYTHIA \cite{Sjostrand:2006za} interfaced to
PHOTOS~\cite{Golonka:2005pn} for final-state radiation.  The
\hwwlnln\ events produced by gluon fusion or vector boson fusion are
modelled using the
MC@NLO\cite{Frixione:2002ik,Frixione:2003ei} and
SHERPA\cite{Gleisberg:2008ta} Monte Carlo generators, respectively.
\hwwlnqq\ is modelled using PYTHIA for the gluon fusion and
HERWIG~\cite{Corcella:2002jc} for vector boson fusion.  The
$\gamma\gamma$ signal is 
simulated with MC@NLO, HERWIG and PYTHIA for the gluon fusion,
vector boson fusion and
associated production processes respectively.

For background sample generation, the PYTHIA,
ALPGEN~\cite{Mangano:2002ea}, MC@NLO, 
MADGRAPH~\cite{Alwall:2007st}, SHERPA  and HERWIG packages  are employed.

All Monte Carlo samples are processed through a complete simulation of the
ATLAS detector~\cite{atlas-sim} using the GEANT
programme~\cite{Agostinelli:2002hh}. 

\section{Search for \hgg}
\label{sec:hgamgam}

The search for the Higgs boson in the $\gamma \gamma$ decay mode is
described below; further details can be found in Ref.~\cite{hgg}. The
event selection requires the 
presence of at least two identified photons~\cite{Aad:2010sp},
including converted  photons,
isolated from any other activity
in the calorimeter. The leading and the sub-leading photons are
required to have transverse momenta above 40~GeV and 25~GeV,
respectively. The directions of the photons are measured using
the position determined in the first sampling of the electromagnetic
calorimeter and that  of the reconstructed  primary vertex. 
The  di-photon invariant mass spectrum is used to search for a peak
above the background contributions. 

\begin{figure}[!htbp]
\begin{center}
\includegraphics[width=0.47\textwidth]{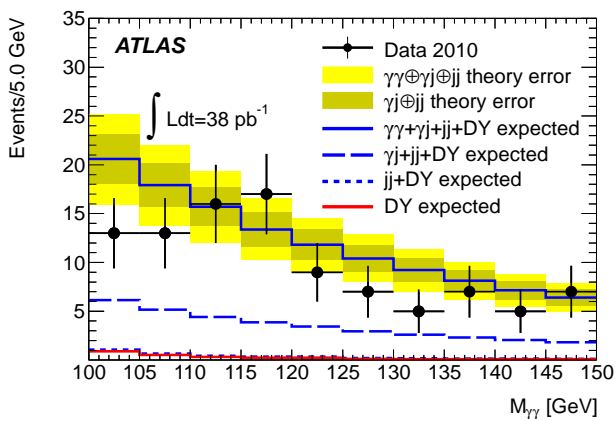}
\end{center}
\caption{Distribution of the di-photon invariant mass for the 99
  events from data passing
all event selection criteria in the \hgg\ search and for the Monte
Carlo prediction.  
The overall uncertainty on the expected total yield is illustrated by the yellow band.
The uncertainty due to the reducible background is also shown (dark yellow band).  The 
predictions for the main components of the background (di-photon,
photon-jet, jet-jet  
and Drell-Yan) are also illustrated.
}

\label{fig:GamGamData}
\end{figure}

The main background processes in the \hgg\ search 
arise from the production of two isolated prompt
photons ($\gamma\gamma$) and  from
fake photons in photon-jet ($\gamma j$) and di-jet ($jj$) events. Fake photons can originate from 
jets in which a leading $\pi^0$ or $\eta$ meson from the 
 quark or gluon fragmentation is reconstructed as a single  isolated
photon. Each of these background 
contributions has been estimated from sideband control samples in the data.  
The backgroud from Drell-Yan events, 
 $Z/\gamma^{*}\to ee$, where the
electrons are mistakenly identified as photons, is estimated from
studies of the $Z$ boson mass peak and extrapolated to the signal
region.
The total number of estimated background events is constrained to be
the observed number.
The di-photon invariant 
mass distribution for the events passing the full selection is
shown in Fig.~\ref{fig:GamGamData}. The full-width at half maximum of a
     signal with \mh=120~\GeV\ would be 4.2~\GeV.

The expected signal yield, summing all production processes, and estimated background composition 
for a total integrated luminosity of 38~\ipb\ are
summarised in \tref{tab:hggyields}. A total of 99 events passing all
 selection criteria are observed in data in the di-photon mass
range from 100~GeV to 150~GeV.  
The background in this region is modelled by fitting an
exponential function to the data. The signal peak is modelled by a 
Gaussian core portion and a power-law low-end
tail~\cite{crystalball}.  Tails in the
signal resolution are modelled by a wide
Gaussian component of small amplitude.
No significant excess of events over the continuous background is found for
any Higgs boson mass. The systematic uncertainty
on the total signal acceptance is $\pm15$\%, where the dominant
contributions come from  photon  
identification ($\pm11$\%) and  photon isolation
efficiencies ($\pm10$\%).

\begin{table}[!htbp]
\caption{The number of expected and observed events in the
  \hgg\ search in the di-photon 
mass range from 100~GeV to 150~GeV for an integrated luminosity of 38~\ipb. 
Also shown is the  composition of the background expected
from Monte Carlo simulation and the division of the observed data 
as discussed in the text  as well as 
the expected number of $H\to\gamma\gamma$ signal events for a 
Higgs boson mass of $\mH=120$~GeV. Total uncertainties are shown in
the middle column while in the
rightmost column 
the statistical and systematic uncertainties, respectively, are given.}
\label{tab:hggyields}
\begin{center}
\begin{tabular}{lcc}
\hline\hline
       & Expected             & Observed or Estimated \\
\hline
Total  &   120 $\pm27$        &   99  \\
\hline
$\gamma\gamma$ &  86$\pm$23   & {75.0$\pm$13.3$^{+2.7}_{-3.6}$} \\
$\gamma j$     & 31$\pm$15    & {19.6$\pm$7.5$\pm$3.9} \\
$jj$           & 1$\pm$1      & {1.5$\pm$0.7$^{+1.8}_{-0.5}$} \\
$Z/\gamma^{*}$  & 2.7$\pm$0.2 & {2.9$\pm$0.1$\pm$0.6} \\
\hline
\hgg\    & \multicolumn{2}{c}{$0.45^{+0.11}_{-0.10}$ ($\mH=120$ GeV)} \\
\hline\hline
\end{tabular}
\end{center}
\end{table}

\section{Search for \hww \label{sec:hww}}

The search for the Higgs boson in the decay channel \hww\ benefits from the large branching 
ratio of the Higgs boson to decay into a pair of $W$ bosons for
masses above $\mH \gtrsim 110$~GeV,  
the sizable $W$ boson decay rates to leptons and the powerful
identification of leptons  
with the ATLAS detector. It offers the greatest
sensitivity of any search channel when the Higgs boson mass is close to twice the $W$ boson mass, 
\mbox{$\mH\sim\! 165$}~\GeV. Two different decay modes of the $W$
bosons are considered:
the \hwwlnln\ channel is pursued for Higgs boson masses in the range
from 120~GeV to 
200~\GeV, and the \hwwlnqq\ decay mode is used for Higgs boson masses in the range from
220~GeV to 600~GeV.  
The analyses are described below and further
details can be found in Refs.~\cite{hwwlnln,hwwlnqq}

\subsection{Search for \hwwlnln \label{sec:HWW-lnln}}

The \hwwlnln\ analysis is performed using a dataset corresponding to
an integrated luminosity of 35~\ipb. Events are selected requiring exactly two isolated leptons with opposite charge. The leading lepton is required to have $\pT > 20$~GeV and 
the sub-leading lepton is required to have \mbox{$\pT > 15$}~GeV. Events are classified into three
channels depending on the lepton flavours: $e\mu$, $ee$ or $\mu\mu$.
If the two leptons are of the same flavour, their invariant mass (\mll) is required to be above 15~GeV
to suppress background from $\Upsilon$ production. To increase the sensitivity, the selections
are then allowed to depend on the Higgs boson mass hypothesis.
For all lepton combinations in the low (high) mass Higgs boson search, \mll\ is required to be 
below 50~(65)~\GeV\ for Higgs boson masses \mbox{$\mH \leq
  170$}~\GeV\ (\mbox{$\mH > 170$}~\GeV) which suppress backgrounds from top-quark production
and $Z$ boson production.
The missing transverse energy in the event is required to be \mbox{$\met > 30$}~GeV.
An upper bound is imposed on the azimuthal angle between the two
leptons, \dphill$< 1.3 \;(1.8)$ radians, taking advantage of the spin
correlations\cite{Dittmar:1996ss} expected in the Higgs boson decay.
The signal region is defined by the transverse mass (\mT)~\cite{Barr:2009mx}:
\begin{linenomath}
\begin{equation}
\label{eq:mT}
m_{\rm T}=\sqrt{(E_{\rm T}^{\ell\ell}+\met)^{2}-({\bf P}_{\rm T}^{\ell\ell}+{\bf P}_{\rm T}^{\rm miss})^{2}},
\end{equation} 
\end{linenomath}
where $E_{\rm T}^{\ell\ell}=\sqrt{({\bf P}_{\rm T}^{\ell\ell})^{2}+m_{\ell\ell}^{2}}$, $|{\bf P}_{\rm T}^{\rm miss}|=\met$ and ${\bf P}_{\rm T}^{\ell\ell}$ is the 
transverse momentum of the dilepton system. The transverse mass is
required to be $0.75\cdot \mH < \mT < \mH$ for the event to be
considered in a given \mH\ range.
Events are also treated separately depending on whether
they have zero jets (0-jet channel) or one jet (1-jet channel)
reconstructed with $|\eta|<4.5$ due to the
differences in background composition and expected signal-to-background ratio. 
To suppress background from top-quark production, 
events in the 1-jet channel are rejected if the jet is identified as coming from a $b$-quark.
Events with two or more jets have been analysed as a separate
channel. However, due to the marginal contribution to the overall
sensitivity given the current total integrated luminosity and the
additional systematic uncertainties,
 this channel is not included in this combination.

The expected background contributions from $WW$, top-quark and \Wjets\ production 
are normalised using dedicated control regions in data as described in
the next sections. Other smaller backgrounds are normalised according 
to their theoretical cross sections. The background from
$Z/\gamma^{*}$+jets production  
is normalised to the
theoretical cross section with a correction factor determined from data.

\subsubsection{The $WW$ background \label{sec:HWW-WWBkg}}

The di-boson $WW$ continuum can be distinguished from the Higgs boson
signal through the kinematic selections. 
 A control region is defined by changing the cut on \mll\  to require
 over 80~\GeV (but not within 10~\GeV\ of the $Z$ boson mass if the leptons are of the same
 flavour)    and removing the selections on $m_{\rm T}$ and \dphill.
The expected ratio of the background contribution 
in the control region and in the signal region is taken from Monte Carlo simulation. The three main sources of 
systematic uncertainty affecting this ratio are the theoretical uncertainty on the extrapolation, the jet energy scale 
uncertainty and the limited statistics in the simulated
sample. Uncertainties due to these effects of $\pm6$\% in the 0-jet channel
and $\pm17$\% in the 1-jet channel have been determined.

\subsubsection{The $t\bar{t}$ and single top-quark backgrounds \label{sec:HWW-topBkg}}

Top-quarks, whether from  strong interaction
\ttbar\ production or  weak interaction single top-quark
production, are a copious source of  final states
with one or two $W$ bosons accompanied by one or more jets. Due to
kinematic selection one or more 
of these jets may fail identification, thereby leading to a final
state similar to that from the \hww\ signal. 

The background from top-quark production in the 0-jet channel is
estimated by first removing the jet veto. This gives a sample
dominated by top-quarks, and  the expected contamination from other
processes in the control region is subtracted from the observed 
event yield. Then the probability that top events  pass the jet
veto  
is derived from the measured probability of not reconstructing a jet
in data,  using a sample of top candidates with two leptons, one
b-jet and no other jet.
 The dominant systematic 
uncertainties originate from the limited statistics in data and 
 the jet energy scale. A total uncertainty of $\pm60$\% has been determined for the top-quark background 
estimate in the 0-jet channel.

The top-quark background in the 1-jet channel is normalised using a control region where the veto on jets coming
from $b$-quarks is reversed and the \dphill, \mll\ and \mT\ selections are removed.
An extrapolation factor from the control region to the signal region is estimated from Monte Carlo
simulation. The dominant systematic uncertainties on the top-quark background estimate in the 
1-jet channel are $\pm23$\%  from the theoretical
uncertainties on the extrapolation factor and $\pm22$\%  from the uncertainty on the $b$-tagging efficiency.

\subsubsection{The $W$+jets background \label{sec:HWW-WjetsBkg}}

The production of $W$ bosons accompanied by jets can mimic the \hww\ signal
if one of the jets is mis-identified as an isolated lepton.
The $W$+jets background is normalised using a control region defined by 
relaxing the identification and isolation criteria for one of the two
leptons.
The contribution to the signal region is estimated by multiplying
the rate measured in the control region by the
probability for fake leptons which pass the relaxed identification and
isolation criteria 
to also pass the original lepton selection criteria. 
This misidentification probability is measured in a multi-jet data sample. The major sources of systematic 
uncertainty for the $W$+jets background estimate come from the bias introduced by the jet trigger
threshold used to select the multi-jet events and the residual
difference in kinematics and 
flavour composition of the jets in multi-jet events and in events from \Wjets\ production. 
The total uncertainty on the estimated $W$+jets background is $\pm50$\%.

\subsubsection{The \Zgjets\ background \label{sec:HWW-ZjetsBkg}}

The largest cross section for producing two isolated,
high-\pT\ leptons comes from the 
$Z/\gamma^{*}\to\ell\ell$ process.
The background from \Zgjets\  is significantly reduced by the upper bound on \mll\ and the
requirement of high \met\ in the signal region. To correct for 
potential mis-modelling of the distribution  of \met\ at high values, a
correction factor is derived from 
the observed difference between the fraction of events passing the \mbox{$\met > 30$}~GeV selection
in data and Monte Carlo simulation for events with \mll\ within 10 GeV
of the  $Z$ boson mass\cite{Nakamura:2010zzi}.
As the discrepancy between
data and Monte Carlo tends to be larger in events with jets, the
correction factor is larger in the 1-jet channel than in the 0-jet channel.
The flavours of the two leptons in the event also impact the magnitude
of the correction factor, 
since any discrepancies between data and simulation have different sources.
In the 1-jet channel, the correction factors are found to be
$1.2\pm0.4\pm0.1$\footnote{When two errors are quoted the first is
  statistical and the second systematic.} in the
$ee$ analysis and $2.4\pm0.5\pm0.2$ in the $\mu\mu$ analysis.
Under the assumption that the same correction factors apply to events below the upper bound on \mll,
the expected \Zgjets\ background is obtained from the Monte Carlo simulation normalised to the product
of the theoretical cross section and the correction factors.

\subsubsection{Results for the \hwwlnln\ search \label{sec:HWW-Results}}

\begin{figure}[htp]
\begin{center}
\subfigure[0-jet channel]{\label{fig:HWW-mT0j}\includegraphics[width=0.49\textwidth]{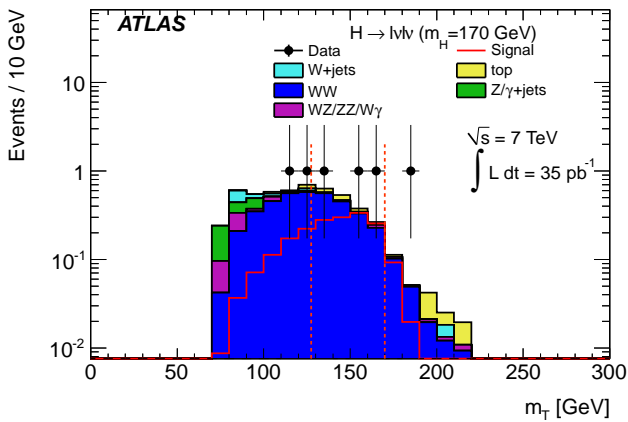}}
\subfigure[1-jet channel]{\label{fig:HWW-mT1j}\includegraphics[width=0.49\textwidth]{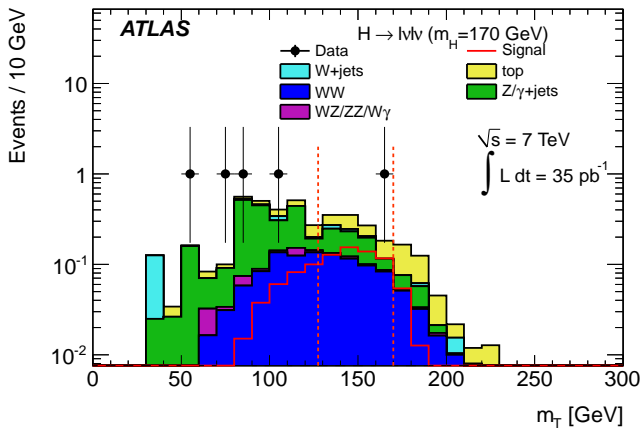}}
\caption{Distributions of the transverse mass \mT\ in the 0-jet channel (a) and 1-jet channel (b) 
for the \hwwlnln\ search
after all selections except the transverse mass cut for the combined
$e\mu$, $ee$ and $\mu\mu$ channels.  The error bars reflect Poisson asymmetric errors.
A Higgs boson signal is shown for $\mH=170$~GeV. 
The selections applied for $\mH=170$~GeV are indicated by the two vertical dotted lines.
\label{fig:HWW-mT} }
\end{center}
\end{figure}

The expected and observed numbers of events in the \hww\ analysis for a Higgs boson mass of 170 GeV are
shown in \tref{tab:nevents170}. Three events in total are observed in the 0-jet channel for the combined
$ee$, $e\mu$ and $\mu\mu$ final states, compared to an expected number of events from background sources
only of $1.70\pm0.12\pm0.17$. More events are expected in the $\mu\mu$ channel compared to the 
$ee$ channel due to different lepton identification efficiencies for electrons and muons.
In the 1-jet channel, one event is observed in the data compared
to a total number of expected events from background sources of $1.26\pm0.13\pm0.23$.
The observed \mT\ distributions in data after all selections except the transverse mass 
cut for the combined $e\mu$, $ee$ and $\mu\mu$ channels are compared to the
expected distributions from simulated events in Fig.~\ref{fig:HWW-mT}.

\begin{table*}[htbp]
\caption{Numbers of expected signal ($\mH=170$~GeV) and background events and the observed numbers of 
events in the data passing all selections in the \hwwlnln\ search. 
The dataset used in this analysis corresponds to an integrated luminosity of 35~\ipb.
The uncertainties shown are
the statistical and systematic uncertainties respectively.
\label{tab:nevents170} }
\begin{center}
\begin{tabular}{lccc}
\hline\hline
 \multicolumn{4}{c}{\bf 0-jet channel} \\
\hline
 & $e\mu$ & $ee$ & $\mu\mu$ \\
\hline
$WW$ & $0.71\pm0.05\pm0.06$ & $0.20\pm0.03\pm0.02$ & $0.53\pm0.02\pm0.05$ \\
\ttbar\ and single top & $0.09\pm0.05\pm0.06$ &$0.03\pm0.01\pm0.02$ &$0.08\pm0.04\pm0.06$  \\
$WZ/ZZ/W\gamma$ & $0.020\pm0.001\pm0.001$ & $0 (<0.001)\pm0$ & $0.010\pm0.001\pm0.001$ \\
\Zgjets & $0 (<0.001)\pm0$ & $0 (<0.001)\pm0$ & $0(<0.002)\pm0$ \\
$W$+jets & $0.01\pm0.01\pm0.01$ &$0.02\pm0.01\pm0.01$ & $0\pm0.10\pm0.01$ \\
\hline
Total Background & $0.83\pm0.07\pm0.13$ & $0.25\pm0.08\pm0.04$ & $0.62\pm0.05\pm0.10$ \\
\hline
\hwwlnln & $0.62\pm0.01\pm0.18$ & $0.20\pm0.01\pm0.07$ & $0.44\pm0.01\pm0.12$ \\
\hline
Observed & 1 & 1 & 1 \\
\hline
\hline
\multicolumn{4}{c}{\bf 1-jet channel} \\
\hline
 & $e\mu$ & $ee$ & $\mu\mu$ \\
\hline
$WW$ & $0.18\pm0.03\pm0.03$ & $0.05\pm0.02\pm0.01$ & $0.16\pm0.03\pm0.02$ \\
\ttbar\ and single top & $0.26\pm0.07\pm0.11$ & $0.10\pm0.02\pm0.04$ & $0.15\pm0.04\pm0.07$ \\
$WZ/ZZ/W\gamma$ & $0.01\pm0.001\pm0.001$ & $0(<0.001)\pm0$  & $0(<0.001)\pm0$ \\
\Zgjets & $0(<0.01)\pm0$ & $0.05\pm0.02\pm0.02$ & $0.25\pm0.08\pm0.05$ \\
$W$+jets & $0.02\pm0.02\pm0.01$ & $0.03\pm0.20\pm0.01$ & $0\pm0.10\pm0.01$ \\
\hline
Total Background & $0.47\pm0.08\pm0.16$ & $0.23\pm0.04\pm0.06$ & $0.56\pm0.09\pm0.14$ \\
\hline
\hwwlnln & $0.31\pm0.01\pm0.09$ & $0.08\pm0.01\pm0.03$ & $0.21\pm0.01\pm0.06$ \\
\hline
Observed & 0 & 0 & 1 \\
\hline\hline
\end{tabular}
\end{center}
\end{table*}

\subsection{Search for \hwwlnqq  \label{sec:HWW-lnqq}}

The \hwwlnqq\ analysis uses a dataset corresponding to an 
integrated luminosity of 35~\ipb.
Events are selected requiring exactly one lepton with \mbox{$\pT > 30$}~GeV.
The missing transverse energy in the event is required to be \mbox{$\met > 30$}~GeV.
Events with fewer than two jets are rejected\footnote{In this channel
  the jet \pT\ threshold is raised from 25~\GeV\ to 30~\GeV.}. Events with $\geq 4$ jets are treated as a separate
search channel, which is however not included in the current combination.
The pair of jets with invariant
mass closest to the  $W$ boson mass is considered to be coming from
the $W$ boson and the measured mass must be between 71~\GeV\ and 91~\GeV. The 
event is rejected if any of the jets in the event is identified as coming from a $b$-quark. 
The invariant mass of the Higgs boson candidate, $m_{\ell\nu qq}$,
is reconstructed with a $W$ boson mass constraint on the
lepton-neutrino system giving rise to a quadratic equation.  If there are
two solutions the one corresponding to the lower 
longitudinal momentum  is taken; if complex the real part is used.

The dominant source of background events 
in the \hwwlnqq\ search comes from $W$+jets production. The
contribution from QCD events is 
estimated by fitting  the observed  $\met$ distribution 
as the sum of templates taken from simulation.
\tref{tab:hwwlnqq_yields} shows the expected numbers of signal and
background events in the signal region, as well as the
observation.
 In the channel with only two and no additional jets,
450 events are observed in the data passing all selection criteria compared to an expected yield from background
sources of $450\pm 41$ events. In the channel with one extra jet 263
events are observed,  compared
to an expected number of background events of $224\pm 15$.

\begin{figure}[hbt]
\begin{center}
\subfigure[0 extra jet channel]{\label{fig:HWW-lnqq0jets}\includegraphics[width=0.47\textwidth]{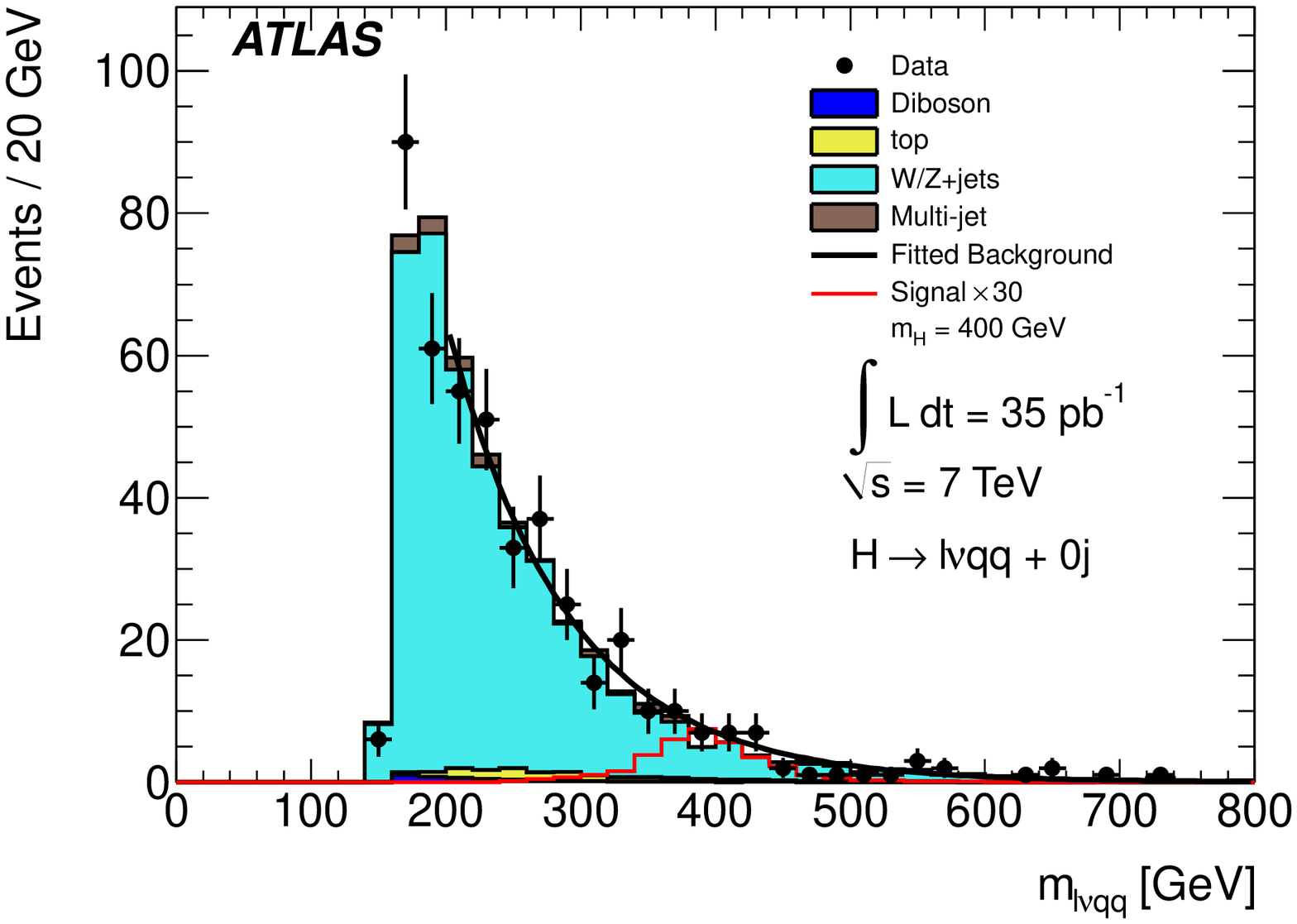}}
\subfigure[1 extra jet channel]{\label{fig:HWW-lnqq1jets}\includegraphics[width=0.47\textwidth]{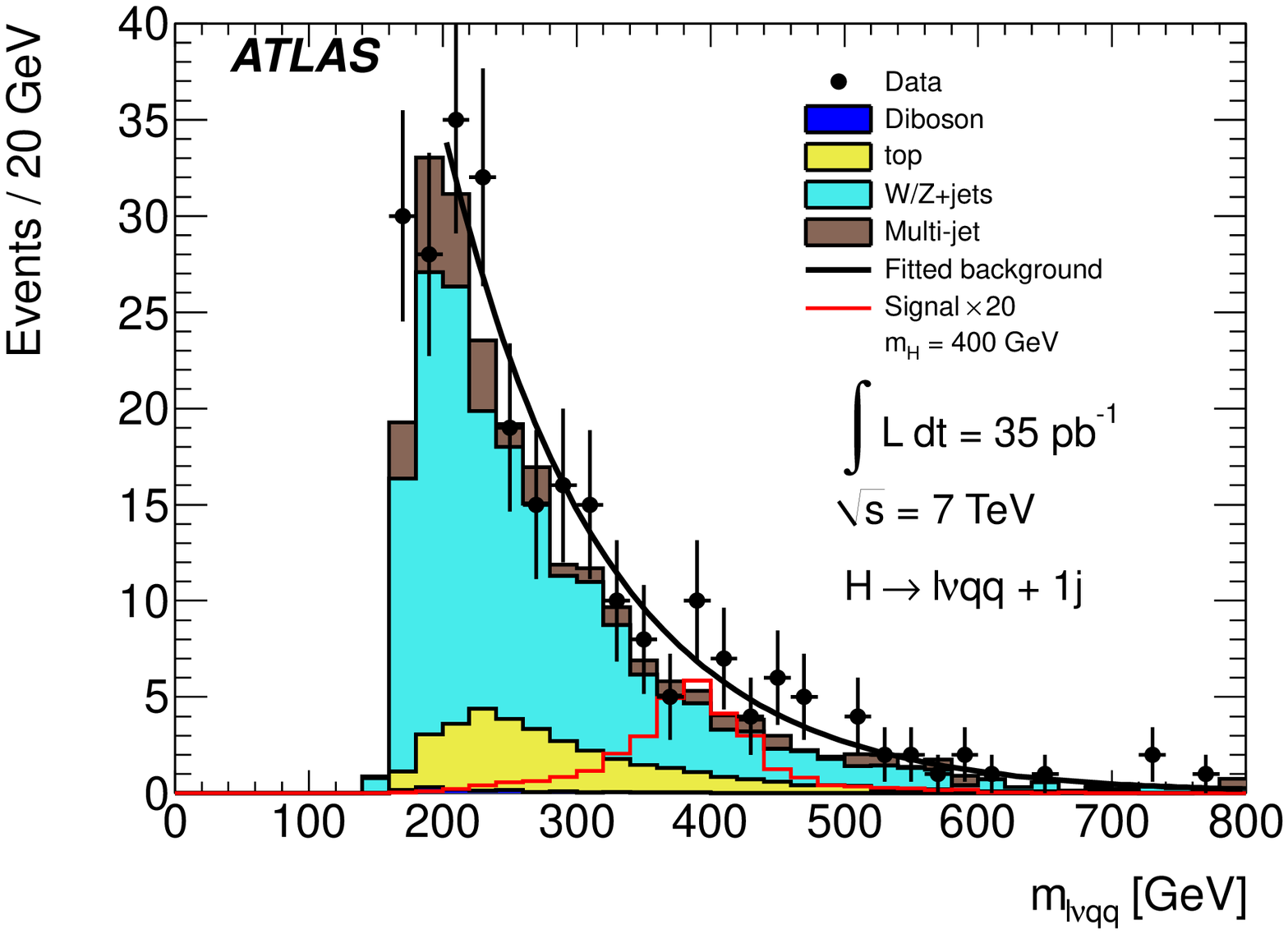}}
\caption{Distributions of the invariant mass $m_{\ell\nu qq}$ for the
  \hwwlnqq\ search after the application of all selection criteria and
  the W-mass constrained fit. The  background fit is   shown  as a
  continuous line. In (a) 
   no extra jets are allowed and in (b) one additional jet  is
   required. The Higgs boson signal is shown for $\mH=400$~GeV and the
   expected yield 
is scaled up by a factor of 30 for illustration purposes.
\label{fig:HWW-lnqq} }
\end{center}
\end{figure}

The distributions of the invariant 
mass for the Higgs boson candidates in data are compared
to the expected distributions from simulated events in
Fig.~\ref{fig:HWW-lnqq}.
The  $m_{\ell\nu qq}$  background spectrum is modelled with a falling
exponential function. The impact of the functional form has been
investigated by replacing the single exponential with a double
exponential, by histograms taken from simulation, and by a mixture of both
methods without significant change in the results.
It should be noted that  the limit extraction is made using the
exponential fit, not 
by comparison with the simulated background.

\begin{table*}[tbp]
\caption{Numbers of expected signal ($\mH=400$~GeV) and background events and the observed numbers of 
events in the data passing all selections in the \hwwlnqq\ search. The dataset used corresponds to an
integrated luminosity of 35~pb$^{-1}$. The quoted uncertainties are combinations of the statistical and
systematic uncertainties. }
\label{tab:hwwlnqq_yields}
\begin{center}
\begin{tabular}{lcccc}
\hline\hline
 $m_H$~=~400~GeV      & \multicolumn{2}{c}{$H+0$-jets} & \multicolumn{2}{c}{$H+1$-jet} \\
\hline
              & $e\nu qq$   & $\mu\nu qq$& $ e\nu qq$  & $\mu\nu qq$ \\
\hline
$W/Z$+jets    & $157 \pm 22$ & $259 \pm 34$ & $39.1 \pm 6.2$ & $119 \pm 12$ \\
Multi-jet     & $11.1 \pm 1.6$ & $4.5 \pm 0.6$ & $17.7 \pm 2.8$ & $13.3 \pm 1.3$ \\
Top           & $5.3 \pm 1.7$ & $7.7 \pm 2.5$ & $15.5 \pm 5.0$ & $18.2 \pm 5.8$\\
Di-boson      & $1.8 \pm 0.3$ & $3.0 \pm 0.4$ & $0.6 \pm 0.1$ & $0.9 \pm 0.1$\\
\hline
Total Background   & $175 \pm 22$ & $275 \pm 34$ & $72.9 \pm 8.4$ & $151 \pm 13$ \\
\hline
\hwwlnqq  & $0.5 \pm 0.2$ & $0.6 \pm 0.2$ & $0.5 \pm 0.2$ & $0.5 \pm 0.2$ \\
\hline
Observed      & 177 & 273 & 87 & 176 \\
\hline\hline
\end{tabular}
\end{center}
\end{table*}

\section{Search for $H \to ZZ^{(*)}$ }
\label{sec:hzz}

Three different $H \to ZZ^{(*)}$ final states are considered here: \hZZllll, \hZZllnn\ 
and \hZZllqq.
In the \hZZllll\ search, the excellent energy and momentum resolutions
of the ATLAS detector for 
electrons and muons lead to a narrow
expected four-lepton invariant mass peak on top of a continuous background. The
dominant background component is the irreducible $\ZZbkg \to \ell\ell\ell\ell$
process. In the low Higgs boson mass region, 
where one of the $Z$ bosons is off-shell and decays into a pair of low
transverse momentum leptons, the reducible backgrounds from
\Zjets\ production and \ttbar\ production are also important. 
For Higgs boson masses above $\mH \gtrsim 200$~\GeV\ both $Z$
bosons are on-shell. In this region the
decay modes \hZZllqq\ and \hZZllnn, which have substantially larger branching
ratios but also larger backgrounds compared to the \hZZllll\ decay,
provide additional  sensitivity.
The analyses of the \hZZllqq\ and \hZZllnn\  channels require that
both $Z$ bosons 
 are on-shell, which limits the contribution
from the reducible backgrounds from \Zjets\ production and
\ttbar\ production. In this paper 
the \hZZllqq\ and  \hZZllnn\ search channels have been used for
Higgs boson masses in the range $200\,\GeV \leq m_H \leq 600\,\GeV$, a
range extending beyond the sensitivities of 
LEP and Tevatron
experiments~\cite{Barate:2003sz,ref:tevhiggs}. Further details of the
three analyses can be found in  Refs.~\cite{hzzllll,hzzllqq}.

\subsection{Search for \hZZllll}
\label{sec:hzzllll}

Candidate events are selected requiring two same-flavour and opposite-charge pairs of leptons.
Muons with $\pT > 7$~GeV and electrons with $\pT > 15$~GeV are considered, 
while at least two out of the four leptons must satisfy $\pT > 20$~GeV.
All leptons are required to be well separated from each other,  isolated
from other activity in the tracking detectors and  the calorimeters
and have low track impact parameters  with respect to the primary vertex.
At least one of the lepton pairs is required to have an invariant mass 
within 15~\GeV\ (within 12~\GeV\ if the combined four-lepton
mass is high) of the $Z$ boson mass.
The requirement on the
invariant mass of the second lepton pair varies as a function of the 
Higgs boson candidate mass, $m_{\ell\ell\ell\ell}$. 
The effective Higgs boson candidate mass resolution $\sigma(\mH)$, 
including the intrinsic width 
at the Higgs boson mass hypothesis being tested, is used to define an allowed
range for the reconstructed Higgs boson candidate mass. The latter is required to be within $\pm 5\sigma(\mH)$ of the tested Higgs
boson mass for the event to be considered.

The magnitude of the \ZZbkg\ background is normalised to the measured $Z$ boson cross section
multiplied by the expected ratio of the cross sections
$\sigma_{ZZ}/\sigma_{Z}$ from theoretical
calculations~\cite{mellado_wu}. 
This estimate is independent of the luminosity uncertainty, and the
cross section ratio is less affected by theoretical uncertainties than
the $\sigma_{ZZ}$ cross section  alone.
The total uncertainty on the \ZZbkg\ background estimate is $\pm15$\%.
The reducible \Zjets\ background arises predominantly from $Z$ boson
production in association
with a pair of heavy flavour quarks which decay
semi-leptonically. This background is normalised
using dedicated control regions in data where the lepton identification requirements are relaxed
for the second pair of leptons. The final uncertainty on the
\Zjets\ background  is $\pm20$\%. 
The \ttbar\ background is estimated from Monte Carlo simulation and normalised to its theoretical cross 
section. A total uncertainty of $\pm25$\% is estimated for the \ttbar\ background contribution.

\begin{figure}[!htbp]
\centering
\includegraphics[width=0.47\textwidth]{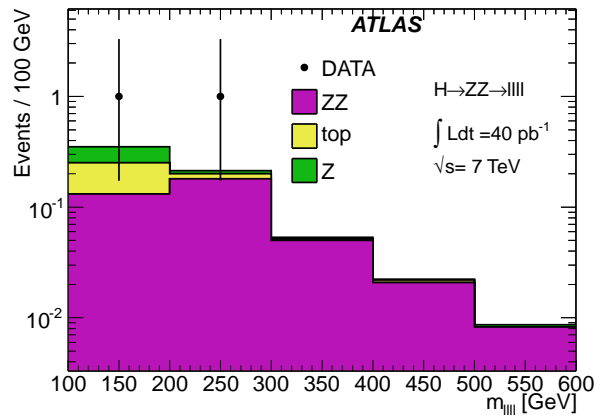}
\caption{
Distribution of $m_{\ell\ell\ell\ell}$ in the \hZZllll\ search 
before  applying the lepton impact parameter and isolation
requirements which remove the two candidates.
The error bars reflect Poisson asymmetric errors.
\label{fig:7TeVlimitHZZ}}
\end{figure}

After the application of all selection criteria, no candidate events remain in data for the 
\hZZllll\ search at any Higgs boson mass. This is consistent with
the small background and signal yields expected with the  integrated
luminosity of 40~\ipb\ used in this analysis.  The results are shown
in \tref{tab:4l-signal-background} for two 
selected Higgs boson masses of $\mH=130$~GeV and $\mH=200$~GeV. The  
distribution of the Higgs boson candidate invariant mass, $m_{\ell\ell\ell\ell}$,
before  applying the lepton impact parameter and isolation
requirements  is shown in Fig.~\ref{fig:7TeVlimitHZZ}.  

\begin{table}[!htbp]
\caption{Expected signal and background event yields in the \hZZllll\ search within $\pm 5\sigma(\mH)$ for
two selected Higgs boson masses.
No events are observed in the data. The dataset used corresponds to a total integrated luminosity of 40~\ipb.
The quoted uncertainties are combinations of the statistical and systematic uncertainties. 
\label{tab:4l-signal-background}}
\begin{center}
\begin{tabular}{lcc}
\hline\hline
\mH~(GeV)           & 130 & 200 \\
\hline
Total background  & 0.010$\pm$0.002  & 0.090$\pm$0.014 \\
\hline
\hZZllll        & 0.015$\pm$0.003 & 0.095$\pm$0.017 \\
\hline
Observed & 0 & 0 \\
\hline\hline
\end{tabular}
\end{center}
\end{table}

\subsection{Search for $\hZZllqq$ \label{sec:hZZllqq}}
Events are selected requiring exactly two same-flavour leptons with an
invariant mass $76\,\GeV < m_{\ell\ell} < 106\,\GeV$ and at least two
jets.
 To reduce  background from top production the missing transverse energy
 is required to be $\met < 50$~GeV.
The two jets in the event with the highest individual \pT\ are required to have
an invariant mass, $m_{jj}$, in the range $70\,\GeV < m_{jj} < 105\,\GeV$. 
Additional background rejection is obtained for  high
mass by using the fact that the final state jets and leptons are
boosted in the directions of the two $Z$ bosons. For Higgs boson searches at 
$\mH \geq 360$~GeV, the two jets are required to have $\pT >
50$~GeV. Furthermore, 
the azimuthal angles between the two jets,  $\Delta\phi_{jj}$, and 
between the two leptons, $\Delta\phi_{\ell\ell}$, must both be less
than $\pi/2$. 

The Higgs boson candidate mass is constructed from the invariant mass of the two leptons and the two jets
in the event, $m_{\ell\ell jj}$. The two jets are constrained to have an invariant mass equal
to the  $Z$ boson mass to improve the Higgs boson candidate mass resolution.

\subsubsection{Background estimates for the $\hZZllqq$ search \label{sec:hZZllqqBkg}}

The dominant background in the \hZZllqq\ search channel is expected to come from \Zjets\ production. Other
significant sources  are \ttbar\ production, multi-jet production and $ZZ/WZ$ production. 
All backgrounds,  except for the multi-jet background, are estimated from Monte Carlo simulation.
For the \Zjets\ and the \ttbar\ backgrounds the predictions from simulation are
compared against data in control samples which are dominated by these
backgrounds.   
The \Zjets\ control region is defined by modifying the $m_{jj}$ selection 
to instead require $40\,\GeV < m_{jj} < 70\,\GeV$ or $105\,\GeV <m_{jj} < 150\,\GeV$. 
The \ttbar\ control region is defined by reversing the 
\met\ selection and modifying the $m_{\ell\ell}$ selection to require $60\,\GeV < m_{\ell\ell} < 76\,\GeV$ or
$106\,\GeV < m_{\ell\ell} < 150\,\GeV$. Both the \Zjets\ and the \ttbar\ background estimates from Monte Carlo simulation
are found to be in good agreement with data in the control
samples. The contribution from \Wjets\ is very small and assumed to be
adequately 
modelled. 
The multi-jet background
in the electron channel is derived from a sample where the electron identification requirements are relaxed. In the muon channel, the multi-jet background is
taken from Monte Carlo after verifying the accuracy of the simulation
using a data sample where the two muons 
in the event are required to have the same charge.

\subsubsection{Results for the $\hZZllqq$ search}

\begin{figure}[htb]
\begin{center}
\includegraphics[width=0.49\textwidth]{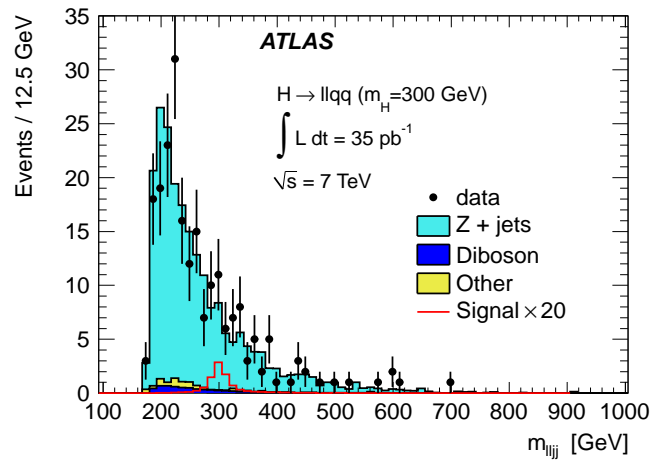}
\caption{Distribution of $m_{\ell\ell jj}$ for events passing all of the selection criteria in
the \hZZllqq\ search. The expected yield for a
Higgs boson with a mass $\mH=300$ GeV is also shown, multiplied by a
factor of 20 for illustrative purposes. The contribution labelled
``Other'' is mostly from top events but includes also QCD multijet production. }
\label{fig:hZZllqqFig}
\end{center}
\end{figure}

The \hZZllqq\ analysis is performed for Higgs boson masses between 200~GeV and 600~GeV in steps of 20~GeV.
Table~\ref{tab:htollqq_yield} summarises the numbers of estimated background events and observed events
in data for the selections below and above $\mH=360$~\GeV. The numbers
of expected  signal events  for
two representative Higgs boson masses are also shown. For the low mass search,
216 events are observed in data passing all selection criteria compared to an expected number of events from background
sources only of $226 \pm 4 \pm 28$ events. The corresponding numbers for the high mass searches are
11 events observed in data compared to an expected yield of $9.9 \pm 0.9 \pm 1.5$ events from background
sources only.
The distribution of the reconstructed Higgs boson candidate mass $m_{\ell\ell jj}$ for the events passing all of the
selection criteria is shown in Fig.~\ref{fig:hZZllqqFig}. 

\begin{table*}[htb]
\caption{Numbers of events estimated as background, observed in data
  and expected from signal in the \hZZllqq\ search for low mass ($\mH <
  360$~GeV) and  high mass  ($\mH \geq 360$~GeV) selections.
The signal, quoted at two mass points, includes small
contributions from $\ell\ell\ell\ell$ and $\ell\ell\nu\nu$ decays.
Electron and muon channels are
combined.  The uncertainties shown are
the statistical and systematic uncertainties, respectively.
}
\label{tab:htollqq_yield}
\begin{center}
\begin{tabular}{ccc}
\hline\hline
  Source   &  low mass selection  &   high mass selection   \\  
\hline
$Z+$jets   &   $214 \pm 4 \pm 27$       &   $9.1 \pm 0.9 \pm 1.4$   \\
$W+$jets   &   $0.33 \pm 0.16 \pm 0.17$   &   $-$   \\
$t\bar{t}$ &   $0.94 \pm 0.09 \pm 0.25$   &   $0.08 \pm 0.02 \pm 0.03$   \\
Multi-jet   &   $3.81 \pm 0.65 \pm 1.91$   &   $0.11 \pm 0.11 \pm 0.06$   \\
$ZZ$       &   $3.80 \pm 0.10 \pm 0.73$   &   $0.30 \pm 0.03 \pm 0.06$   \\
$WZ$       &   $2.83 \pm 0.05 \pm 0.88$   &   $0.29 \pm 0.02 \pm 0.10$   \\
\hline
Total background &   $226 \pm 4 \pm 28$   &   $9.9 \pm 0.9 \pm 1.5$  \\
\hline
\hZZllqq   & $0.60 \pm 0.01 \pm 0.12$ ($\mH=200$ GeV) & $0.24 \pm (<0.001) \pm 0.05$ ($\mH=400$ GeV) \\
\hline
Observed       &   216               &   11             \\ 
\hline\hline
\end{tabular}
\end{center}
\end{table*}

\subsection{Search for $\hZZllnn$ \label{sec:hZZllnn}}
The \hZZllnn\ final state is characterised by two charged leptons and large \met. 
Events are selected by requiring exactly two  leptons of the same
flavour with  an
invariant mass $76\,\GeV < m_{\ell\ell} < 106\,\GeV$.
Events are rejected if any jet  is identified as coming from a $b$-quark.
The  selection has been optimised separately for searches at low ($\mH < 280$~GeV) and high 
($\mH \geq 280$~GeV) values of the Higgs boson mass. Events are required to have $\met>66\;(82)\,$~GeV and 
$\Delta\phi_{\ell\ell}<2.64\;(2.25)$ radians for the low~(high) mass region.
For the low mass region $\Delta\phi_{\ell\ell}>1$ radian is also required.  
The Higgs boson candidate transverse mass is 
obtained from the invariant mass of the two leptons and the
missing transverse energy.

\subsubsection{Background estimates for the \hZZllnn\ search}
A major  background in the \hZZllnn\ search channel comes  from
di-boson production and  is estimated from Monte Carlo simulation.
Background contributions from \ttbar\
and $W$+jets production are also obtained from Monte Carlo simulation, and the estimated yields
are verified by comparing with the number of observed events in dedicated control samples
in the data. Both the \ttbar\ and the $W$+jets control regions are defined by modifying
the $m_{\ell\ell}$ selection to instead require $60\,\GeV < m_{\ell\ell} < 76\,\GeV$ or
$106\,\GeV < m_{\ell\ell} < 150\,\GeV$. The \ttbar\ control region also requires that the events
pass $\met>20$~GeV and that at least one jet is identified as coming from
a $b$-quark. The $W$+jets control region instead requires 
$\met>36$~GeV and that 
no jets in the events are identified as coming from a $b$-quark. The
observed event yields 
in the control regions for \ttbar\ and $W$+jets production are in good
agreement with the predictions  
from the Monte Carlo simulation. The background from
\Zjets\ production is estimated from Monte Carlo simulation after
comparison studies of the  \met\ distribution between Monte Carlo and data.
The multi-jet background in the electron channel is derived from a sample where the electron 
identification requirements are relaxed. 
In the muon channel, the multi-jet background is estimated from a
simulated sample of semi-leptonically
decaying $b$- and $c$-quarks and found to be negligible after the application of the $m_{\ell\ell}$ 
selection. This was verified in data using leptons with identical charges.

\subsubsection{Results for the \hZZllnn\ search}

\begin{figure}[htb]
\begin{center}
\includegraphics[width=0.49\textwidth]{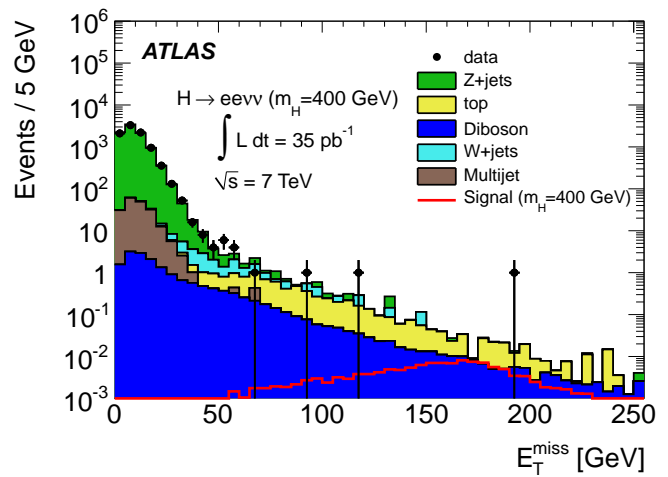}
\caption{Distribution of missing transverse energy in the
  \hZZllnn\ search in the electron channel before vetoing events with low \met. The expected yield for a
Higgs boson with $\mH=400$ GeV is also shown. The  distribution in the
muon channel is similar with four events seen which have
\met\ above 80 GeV.}
\label{fig:hZZllnnFig}
\end{center}
\end{figure}

The \hZZllnn\ analysis is performed for Higgs boson masses between 200~GeV and 600~GeV in steps of 20~GeV.
Table~\ref{tab:htollvv_yield} summarises the numbers of events
observed in the data, the estimated numbers of background 
events and the expected numbers of signal events for two selected \mH\ values.  For the low mass selections,
five events are observed in data  compared to an expected number of events from background
sources only of $5.8 \pm 0.5 \pm 1.3$. The corresponding results for the high mass selections are
five events observed in data compared to an expected yield of $3.5 \pm 0.4 \pm 0.8$ events from background
sources only. In addition to the \hZZllnn\
decays, several other Higgs boson channels give a non-negligible
contribution to the total expected signal yield.  
In particular, \hwwlnln\ decays can lead to final states that are very
similar to
\hZZllnn\ decays. They 
are found to contribute significantly to the signal yield at low
\mH\ values. The expected number of  
events from \hwwlnln\ decays relative to that from \hZZllnn\ decays is 76\% 
for $\mH=200$~GeV and 9\% for $\mH=300$~GeV. The kinematic selections
prevent individual candidates from being accepted by both searches.
The \met\ distribution
before vetoing events with low \met\ is shown in Fig.~\ref{fig:hZZllnnFig}.

\begin{table*}[h!tb]
\caption{Numbers of events estimated from background, observed  in
  data and expected from  signal 
 in the \hZZllnn\ search for low mass ($\mH <
  280$~GeV) and  high mass  ($\mH \geq 280$~GeV) selections.
Electron and muon channels are
combined.  The expected signal events include  minor
additional contributions from \hZZllqq\ , \hZZllll\  and  one which
can be large from \hwwlnln.
The uncertainties shown are
the statistical and systematic uncertainties, respectively. 
}
\label{tab:htollvv_yield}
\begin{center}
\begin{tabular}{ccc}
\hline\hline
 Source   & low mass selection   & high mass selection  \\  
\hline
 $Z+$jets   &   $1.09 \pm 0.29 \pm 0.59$       &   $1.01 \pm 0.29 \pm 0.58$   \\
 $W+$jets  &   $1.07 \pm 0.31 \pm 0.64$   &   $0.41 \pm 0.19 \pm 0.22$   \\
 $t\bar{t}$  &   $1.90 \pm 0.10 \pm 0.63$   &   $0.91 \pm 0.07 \pm 0.31$   \\
 Multi-jet    &   $0.11 \pm 0.11 \pm 0.06$   &   $-$   \\
 $ZZ$        &   $0.58 \pm 0.01 \pm 0.11$   &   $0.51 \pm 0.01 \pm 0.10$   \\
 $WZ$        &   $0.57 \pm 0.01 \pm 0.10$   &   $0.45 \pm 0.01 \pm 0.09$   \\ 
 $WW$       &   $0.43 \pm 0.02 \pm 0.09$   &   $0.16 \pm 0.01 \pm 0.04$   \\ 
\hline
 Total background &   $5.8 \pm 0.5 \pm 1.3$   &   $3.5 \pm 0.4 \pm 0.8$  \\
\hline 
\hZZllnn      & $0.19 \pm (< 0.001) \pm 0.04$ ($\mH=200$ GeV)   & $0.30 \pm (<0.001) \pm 0.06$ ($\mH=400$ GeV) \\
\hline
 Observed       &   5               &   5             \\ 
\hline\hline
\end{tabular}
\end{center}
\end{table*}

\begin{table*}[htbp]
\caption{Summary of  systematic uncertainties (in percent) of the
  signal yield. 
The correlated systematic uncertainties are given in detail, the
uncorrelated ones are lumped together. 
The uncertainties are evaluated for a Higgs boson mass of 115~\GeV\ for the
\hgg\ channel, 160~\GeV\ for \hwwlnln, 200~\GeV\ for the \hZZllll\ and
400~\GeV\ for the remaining channels. Systematic errors marked with a
dash are neglected. In the three channels, the
impact of the lepton
energy scale and resolution uncertainties on the efficiency  was found
to be negligible, but they can still influence the fit via the signal distributions.
 In the \hwwlnqq\ channel a jet energy scale uncertainty can only decrease
 the efficiency; the resolution uncertainty is negligible in comparison.\label{ta:sys}}
\begin{center}
\begin{tabular}{lcccccc}
\hline \hline
                       &$\gamma\gamma$& \multicolumn{2}{c}{\hww}& \multicolumn{3}{c}{\hzz} \\
                       &              & $\ell\nu \ell\nu$ & $\ell\nu qq$  &
$\ell\ell\ell\ell$ & $\ell\ell\nu\nu$& $\ell\ell qq$ \\

\hline
Luminosity             & $\pm 3.$4    & $\pm 3.4$     & $\pm 3.4$    & $\pm 3.4$    & $\pm 3.4$ & $\pm 3.4$   \\
\hline
e/$\gamma$ efficiency  & $\pm$ 11   & $\pm 8.2$ & $\pm 2.6$&$\pm2.1$&$\pm4.6$&$\pm0.0$ \\
e/$\gamma$ energy scale& -            & $\pm 1.6$ & -          &  -       &$\pm0.5$&$\pm0.2$ \\
e/$\gamma$ resolution  & -            & $\pm 1.6$ & -          &  -       &$\pm0.1$&$\pm0.1$  \\
\hline
 $\mu$ efficiency      &    -         & $\pm 0.5$ & $\pm$1.0 &$\pm0.8$&$\pm0.0$&$\pm2.0$ \\
 $\mu$ energy scale    &    -         & $\pm 4.8$ & -          & -       & $\pm1.2$&~$^{+0.2}_{-2.2}$ \\
 $\mu$ resolution      &      -       & $\pm 1.2$ & -          &  -       &$\pm0.1$&$\pm0.1$ \\
\hline
Jet energy scale       &     -        &  $\pm 3.7$ & -26     & -       &$\pm0.4$&~$^{+2.9}_{-7.0}$ \\
Jet energy resolution  &     -        &   -          &  -        & -       &$\pm0.2$&~$^{+0.0}_{-1.3}$ \\
\hline
 $b$-tag efficiency    &     -        & -            & -         & -       &$\pm0.4$& -\\
\hline
 Uncorrelated          & $\pm$10    &   $\pm$5.0 &   -       &  -       & -       & -\\
\hline\hline
\end{tabular}
\end{center}
\end{table*}

\section{Combination method \label{sec:stat}}

The limit-setting procedure uses the power-constrained profile likelihood method
known as the Power Constrained Limit, PCL~\cite{ref:pcl,cscbook,Cowan:2010st}. 
This method is preferred to the more familiar CL$_s$\cite{Read:2002hq} technique
because the constraint is more transparently defined 
and it has reduced overcoverage resulting in a more precise meaning of the
quoted confidence level. 
 The resulting PCL median limits have been found to be  around 20\%
 tighter than those obtained with the CL$_s$ method in several Higgs searches.
The application of  
the PCL method to each of the individual Higgs boson search channels
is described in Refs.~\cite{hgg,hwwlnln,hwwlnqq,hzzllll,hzzllqq}.
A similar procedure is used here. 
The individual analyses are combined by maximising the product of
the likelihood functions for each channel and computing a likelihood ratio. A 
single signal normalisation parameter $\mu$ is used for all analyses,
where $\mu$ is the ratio of the hypothesised cross section to the  
expected \SM\ cross section.

Each channel has sources of systematic uncertainty, some of which are
common with other channels. Table~\ref{ta:sys} lists the common sources of systematic uncertainties, which
are taken to be 100\% correlated with other channels. Let the
search channels be labelled by $l$ ($l=$ \hgg, \hww, \ldots), 
the background contribution, $j$, to channel $l$ by
$j_l$ and the systematic uncertainties by $i$
($i=$ luminosity, jet energy scale, \ldots). The
relative magnitude of the effect on the Higgs boson signal yield in
channel $l$ due to  systematic uncertainty $i$ is then denoted by
$\epsilon^s_{li}$, and on background contribution $j_l$,
$\epsilon^b_{jli}$. The $\epsilon_{li}$'s are constants; an individual
$\epsilon^s_{li}$ can be zero if the channel in question is not affected
by this source of systematic uncertainty.
A common systematic uncertainty, i,  which is shared between
channels $l$ and $l^{\prime}$ implies that $\epsilon^s_{li}$ and
$\epsilon^s_{l^{\prime}i}$ are both different from zero. If a systematic source $i$ is shared
between the signal in channel $l$ and background contribution $j_l$ then both
$\epsilon^s_{li}$ and $\epsilon^b_{jli}$ are non-zero. For each source of systematic
uncertainty $i$ there is a corresponding nuisance parameter $\delta_i$ and an associated 
auxiliary measurement $m_i$ on a control sample (e.g. sidebands in a
mass spectrum) that is used to constrain 
the parameter.  The $\delta_i$ and $m_i$ are scaled so that
$\delta_i=0$ corresponds to the nominal expectation and $\delta_i=\pm
1$ corresponds to the $\pm 1\sigma$ variations of the source.
When constructing ensembles for statistical evaluation, each
$m_i$ is sampled according to $G(m_i|\delta_i,1)$, the standard normal
distribution.
Using this notation, the total number of expected events in the signal
region for channel $l$ is given by:
\begin{linenomath}
\begin{eqnarray}
\label{eq:muT}
\nonumber   N^{exp}_l & = & \mu L \;\sigma_l \prod_i(1+\epsilon^s_{li}\delta_i) + \\ 
 &  & \sum_{j} b_{jl}\prod_i(1+\epsilon^b_{j_li}\delta_i)
\end{eqnarray}
\end{linenomath}
for luminosity $L$, \SM\ cross sections $\sigma_l$ (including
efficiencies and acceptances), and expected backgrounds
$b_{jl}$. Background estimates $b_{jl}$ may come  either from Monte Carlo simulations or from control regions in which the expected number of events, $\bar{n}_{jl}$, is proportional to the expected background, via $b_{jl} = \alpha_{jl}\bar{n}_{jl}$.
Given the number of observed events in the signal region $N^{obs}_l$, the likelihood function can be written as:
\begin{linenomath}
\begin{eqnarray}
\label{eq:SingleLikelihood}
\nonumber \lhood_l & = & \textrm{Pois}(N^{obs}_l | N^{exp}_l) \;\prod_{j_l} \textrm{Pois} (n_{jl}|\bar{n}_{jl}) 
 \prod_{i} G(m_i|\delta_i,1)
\end{eqnarray}
\end{linenomath}
where $n_{jl}$ are the observed numbers of background events in the control regions  and 
$\textrm{Pois}(x|y)$ is the Poisson probability of observing $x$ events given an expectation $y$.

The combined likelihood is given by the product of the individual likelihoods for each channel:
\begin{linenomath}
\begin{eqnarray}
\label{eq:likelihood}
\nonumber \lhood  & = & \prod_l \lhood_l,
\end{eqnarray}
\end{linenomath}
where $l$ is implicitly an index over the individual histogram bins within the channels that used a binned distribution of a discriminating variable.

The profile likelihood ratio
\begin{linenomath}
\begin{equation}
\label{eq:lambda}
\tilde\lambda(\mu) = \left\{ 
\begin{array}{l l}
 \frac{\lhood(\mu,\hat{\hat{\theta}}(\mu))}{\lhood(\hat{\mu},\hat{\theta})},  & \hat{\mu} \ge 0,\\
 \frac{\lhood(\mu,\hat{\hat{\theta}}(\mu))}{\mathcal{L}(0,\hat{\hat{\theta}}(0))},  & \hat{\mu} < 0,\\ \end{array} \right. 
\end{equation}
\end{linenomath}
is computed by maximising the likelihood function twice: in the numerator $\mu$, the ratio of the hypothesised cross section to the expected Standard Model cross section, is restricted to a
particular value and in the denominator 
$\mu$ is allowed to float. The set of all nuisance parameters $\delta_i$ and $\bar{n}_{jl}$ is denoted
$\theta$. The maximum likelihood estimates of  $\mu$  and $\theta$ are denoted $\hat{\mu}$ and 
$\hat{\theta}$, while $\hat{\hat{\theta}}(\mu)$ 
denotes the conditional maximum likelihood estimate of all nuisance
parameters with $\mu$ fixed. 
In this analysis the range of $\mu$ is restricted to 
the physically meaningful regime, i.e. it is not allowed to be negative. 
The test statistic 
$\tilde q_{\mu}$ is defined to be 
\begin{linenomath}
\begin{eqnarray}
\label{eq:lambda2}
\nonumber \tilde q_{\mu} & = & \left\{ 
\begin{array}{l l}
 -2\ln\tilde\lambda(\mu), & \hat{\mu} \le \mu, \\
 0,  & \hat{\mu} > \mu, \\ \end{array} \right. \\
& = & 
 \left\{
\begin{array}{l l l}
 -2\ln\frac{\lhood(\mu,\hat{\hat{\theta}}(\mu))}{\mathcal{L}(0,\hat{\hat{\theta}}(0))},  & \hat{\mu} < 0, \\
 -2\ln\frac{\lhood(\mu,\hat{\hat{\theta}}(\mu))}{\lhood(\hat{\mu},\hat{\theta})},  & 0 \le \hat{\mu} \le \mu,\\
 0, & \hat{\mu} > \mu. \\ \end{array} \right. 
\end{eqnarray}
\end{linenomath}
Monte Carlo pseudo-experiments are generated to construct the probability density function
$f(\tilde q_{\mu}|\mu,\bf{\hat{\hat{\theta}}}(\mu))$ under an assumed
signal strength $\mu$, giving a $p$-value 
\begin{linenomath}
\begin{equation}
\label{eq:pmu}
 p_{\mu}=\int_{\tilde q_{\mu,obs}}^{\infty} \! f(\tilde q_{\mu}|\mu,\bf{\hat{\hat{\theta}}}(\mu)) \, \mathrm{d\tilde q_{\mu}}. \\
\end{equation}
\end{linenomath}
To find the upper limit on $\mu$ at 95\% confidence level, $\mu_{up}$, $\mu$ is varied to find 
$p_{\mu_{up}} = 5\%$. Similarly, background-only Monte 
Carlo pseudo-experiments are used to find the median $\mu_{med}$ 
along with the $\pm 1\sigma$ and $+2\sigma$ bands expected in the
absence of a signal. 
The procedure so far can be referred to as a CL$_{sb}$  limit. 
To protect against excluding the (signal) null hypothesis in cases of
downward fluctuations of the background, the observed
limit is not allowed to fluctuate below the $-1\sigma$ expected
limit. This is equivalent 
to restricting the interval to cases in which the statistical power of
the test of $\mu$ against the alternative $\mu=0$ is at least
16\%. This is referred to as  a Power Constrained Limit. 
If the observed limit fluctuates below the 16\% power, the quoted
limit is $\mu_{med}-1\sigma$.

\section{Systematic uncertainties in the combination} 
\label{sec:systematics}

The systematic uncertainty related to the luminosity is
\lumiuncertainty\ and is fully correlated
among all channels. It affects  
background estimates that are normalised to their theoretical cross sections;
for most channels this is only true for backgrounds that are known to be
small.  In the \hZZllnn\ and \hZZllqq\ channels
major backgrounds are normalised to their theoretical cross sections,
but in the latter case this is only done after comparing with control regions.

Sources of systematic uncertainty related to the event
reconstruction are correlated between all the Higgs boson search channels.  
The uncertainty on the efficiency to reconstruct
electrons varies between 2.5\% (central high-\pT\ electrons) and 16\% (\pT\ near
15~\GeV, the lowest value used here) 
but it is assumed to be completely correlated. 
For muons the efficiency uncertainty ranges between
0.4\% and 2\%. The jet systematic errors are typically larger for the channels
where  jets are explicitly required. They are dominated by the jet
energy scale as the  resolution effects tend to partially cancel and
the \MET\ uncertainties  are largely by-products of the uncertainties
already discussed.

The effect on the signal yield  in each channel of the major
sources of systematic uncertainty  
is summarised in Table~\ref{ta:sys}. Uncertainties are treated as
either uncorrelated 
or 100$\%$ correlated among channels. The largest uncorrelated errors are
 photon isolation in \hgg\ and jet rates in \hwwlnln; the latter is in
 principle correlated with the \hwwlnqq\ channel but these channels are
 never used in the same mass region.
Most backgrounds have been estimated by means of independent control samples;
these estimates are assumed to be uncorrelated between the channels. 

Systematic uncertainties on the signal shape are accounted for in the
\hgg,  \hZZllqq\ and \hZZllnn\ channels 
by considering three possible  distributions and interpolating between
them.
Small signal shape systematic uncertainties in the
\hZZllll\ and \hwwlnqq\ channels are neglected.
For $\mH \geq 200$~GeV 
 the correlations in the shape systematics are
taken into account and are treated as correlated with the signal
normalisation uncertainties.

The width of the Higgs boson signal at high mass is taken from the
PYTHIA Monte Carlo~\cite{Sjostrand:2006za}. This underestimates the
width and the
accepted cross section  is conservatively scaled down
by the ratio of the widths given in Ref.~\cite{Djouadi:2005gi}, which
reached a maximum of 8\% at 600~\GeV, in all
plots showing a ratio to the Standard Model.

The systematic uncertainty coming from the total theoretical Higgs
boson cross section is  
not included in the combination and is shown separately in the figures as an
uncertainty on the predicted cross section. 

\section{Combination}
\label{sec:combination}

Each Higgs boson search channel is only sensitive for a range of 
Higgs boson masses. The ranges in which the various channels
have been analysed are detailed in Table~\ref{ta:contrib}.
In the \hgg, \hZZllnn\ and \hZZllqq\ channels, the final result is extracted from
a fit of signal plus background contributions
to the observed Higgs boson candidate mass distributions. In all other
channels, limits are extracted from
 a comparison of the numbers of observed events in one or more signal
regions to the numbers of estimated background events.

\begin{table}[!htbp]
  \begin{center}
  \caption{The Higgs boson mass regions in which individual search
    channels have been analysed.
\label{ta:contrib}
}
\vspace*{0.2cm}
    \begin{tabular}{cc}
      \hline
      \hline
Mode & Mass range, \GeV \\
      \hline
\hgg     &  110 -- 140\\
\hwwlnln &  120 -- 200\\
\hwwlnqq &  220 -- 600\\
\hZZllll &  120 -- 600\\
\hZZllnn &  200 -- 600 \\
\hZZllqq &  200 -- 600 \\
      \hline \hline
    \end{tabular}
  \end{center}
\end{table}

\begin{figure*}[!htpb]
\centering  
\mbox{\hspace*{0.1cm}{\includegraphics[width=0.85\textwidth]{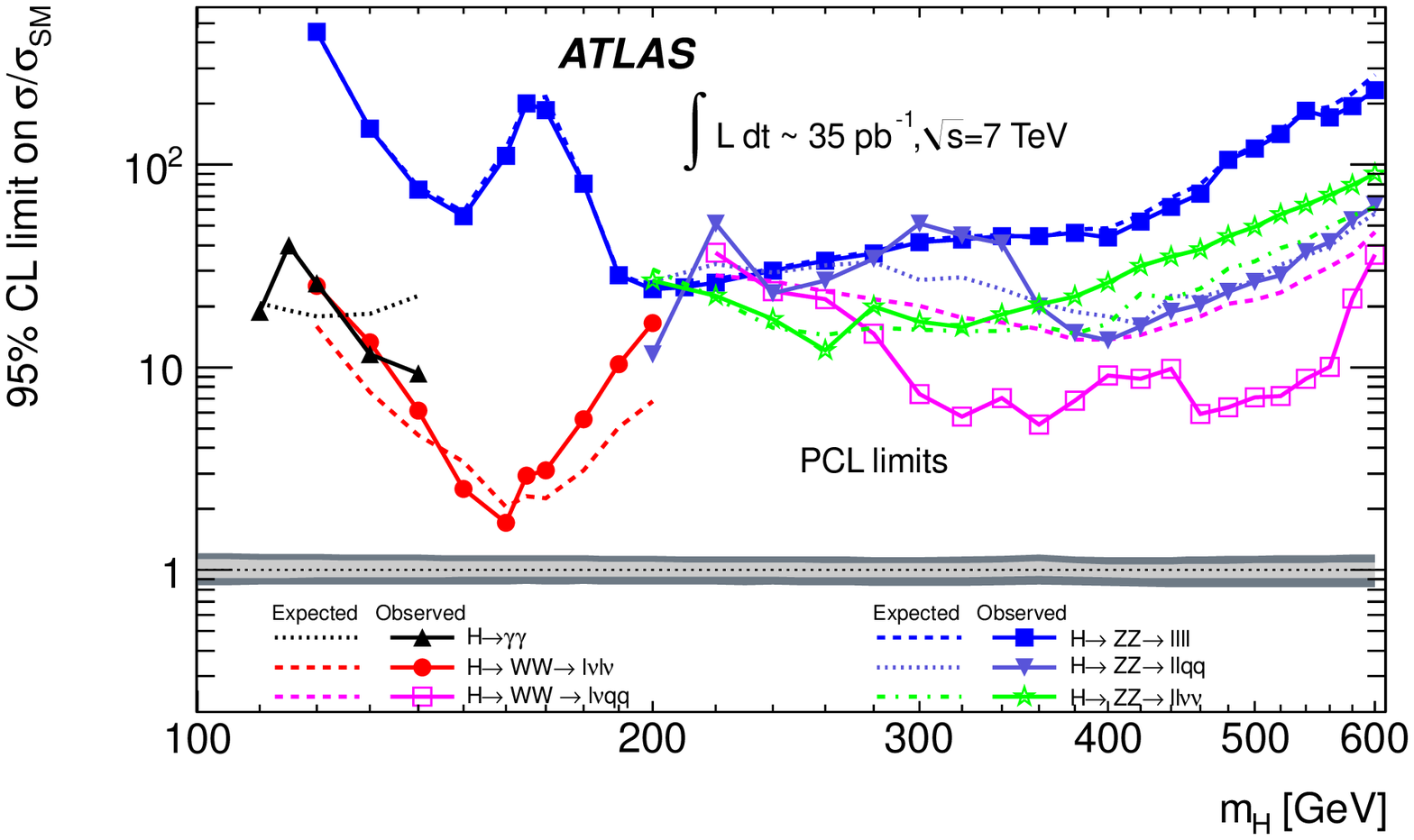}}}
\caption{The expected and observed cross section limits, normalized to
  the \SM\ cross section, as a function of the Higgs boson mass for
  the individual search channels.
The visually most apparent difference between expected and observed
is in the \hwwlnqq\ channel, which has a  deficit approaching one sigma
both at 320~\GeV\ and 480~\GeV.
These results use the profile likelihood method with a power
constraint (PCL).
The limits are calculated at the masses marked with symbols.
 The lines between the points are to guide the eye.
The grey horizontal bands  show the uncertainty on the \SM\ cross section
prediction, with the inner region highlighting the contribution of QCD
scale uncertainties.
\label{fig:limit-inputs}
}
\end{figure*}

The individual channels  are
  shown in Fig.~\ref{fig:limit-inputs} in terms of the observed and the
  expected upper limits on $\sigma/\sigma_{\mathrm {SM}}$ at the 95\%
  confidence level.  The step with which the limit is extracted,
  5-10~\GeV, does not match the \hgg\ resolution which
  has a full-width at half maximum of 4.4~\GeV. However, it
  was established in Ref.~\cite{hgg} that no important fluctuations
  are missed.

\begin{figure*}[!p]
\centering  
\mbox{\hspace*{0.1cm}{\includegraphics[width=0.85\textwidth]{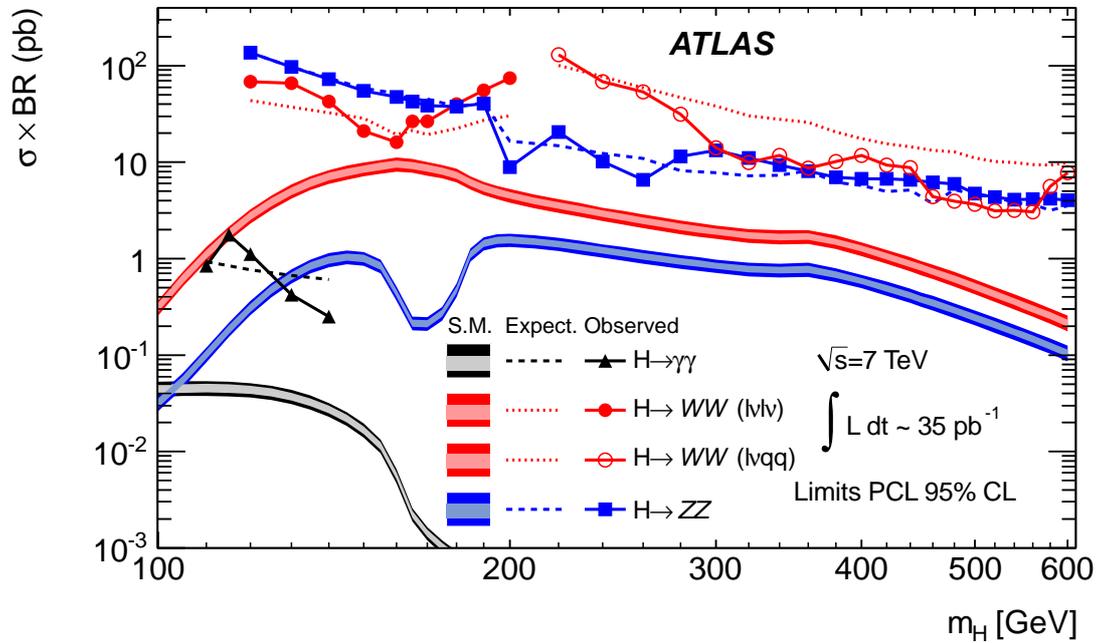}}}
\caption{The expected and observed 95\% PCL limits on the total cross section
  of a particle 
produced like the \SM\ Higgs boson and decaying with the width
predicted by PYTHIA\cite{Sjostrand:2006za} to pairs of bosons:
$\gamma\gamma$, WW or ZZ .
The limits are calculated at the masses marked with symbols.
 The lines between the points are to guide the eye.
The coloured  bands show the cross section
predictions and their uncertainties, with
 the inner region highlighting the contribution of QCD
scale uncertainties.
\label{fig:limit-sig}
}
\end{figure*}

The search channels are grouped by the primary Higgs boson decay mode
searched for, $\gamma\gamma$, WW or ZZ, and the limit on each mode is extracted
in terms of the cross section for the process intended.
Some channels have a contribution from signal modes  other then the
intended one. This is 
only significant for the \hZZllnn\  search, as discussed in
Section~\ref{sec:hZZllnn}, 
and implies that the ZZ limit  assumes the \SM\ ratio between  H to
ZZ and H to WW decays.
In addition, the WW search  requires zero or one  jet and is
essentially designed for a spin-zero object produced largely via gluon fusion. 
The upper limits at 95\% confidence level observed and expected in the
absence of a signal are
compared with the 
cross section expected for a Standard Model Higgs boson in 
Fig.~\ref{fig:limit-sig}. 

\begin{figure*}[!p]
\centering  
\mbox{\hspace*{0.1cm}{\includegraphics[width=0.85\textwidth]{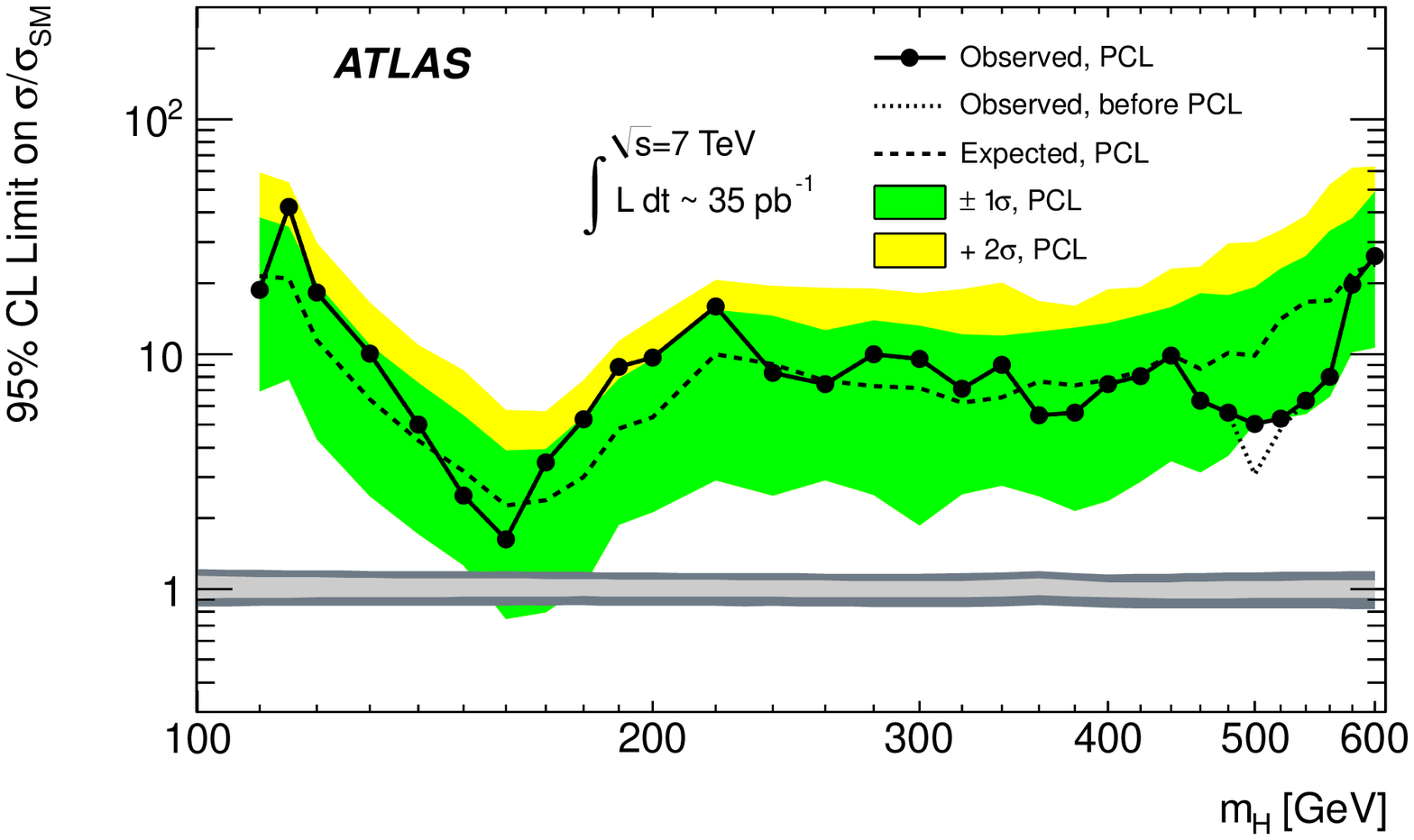}}}
\caption{The expected and observed upper limits on the total cross section divided by the expected \SM\ Higgs boson
cross section. This is a 95\% PCL limit.
The green and yellow bands indicate the range in which  the
limit is expected to  lie in the absence of a signal.
The fine dotted line marks the results obtained using CL$_{sb}$,
and the application of the power constraint gives the solid line.
The limits are calculated at the masses marked with symbols and
 the lines between the points are to guide the eye.
\label{fig:limit-div}
}
\end{figure*}

The combination of all channels is tested and 
the $p_{\mu=0}$ in these fits varies between 7\% and 89\%, which
does not suggest the presence of a signal.
The combination of all channels is
  shown in Fig.~\ref{fig:limit-div} in terms of the observed and the
  expected upper limit at the 95\% confidence level.  
The statistical accuracy of the toy Monte Carlo used to extract the
limits is about 5\% on the observed limits and 7\% on the expected ones,
with somewhat larger variation on the edges of the one and two
$\sigma$ bands.

\begin{figure*}[!p]
\centering  
\mbox{\hspace*{-0.1cm}{\includegraphics[width=0.85\textwidth]{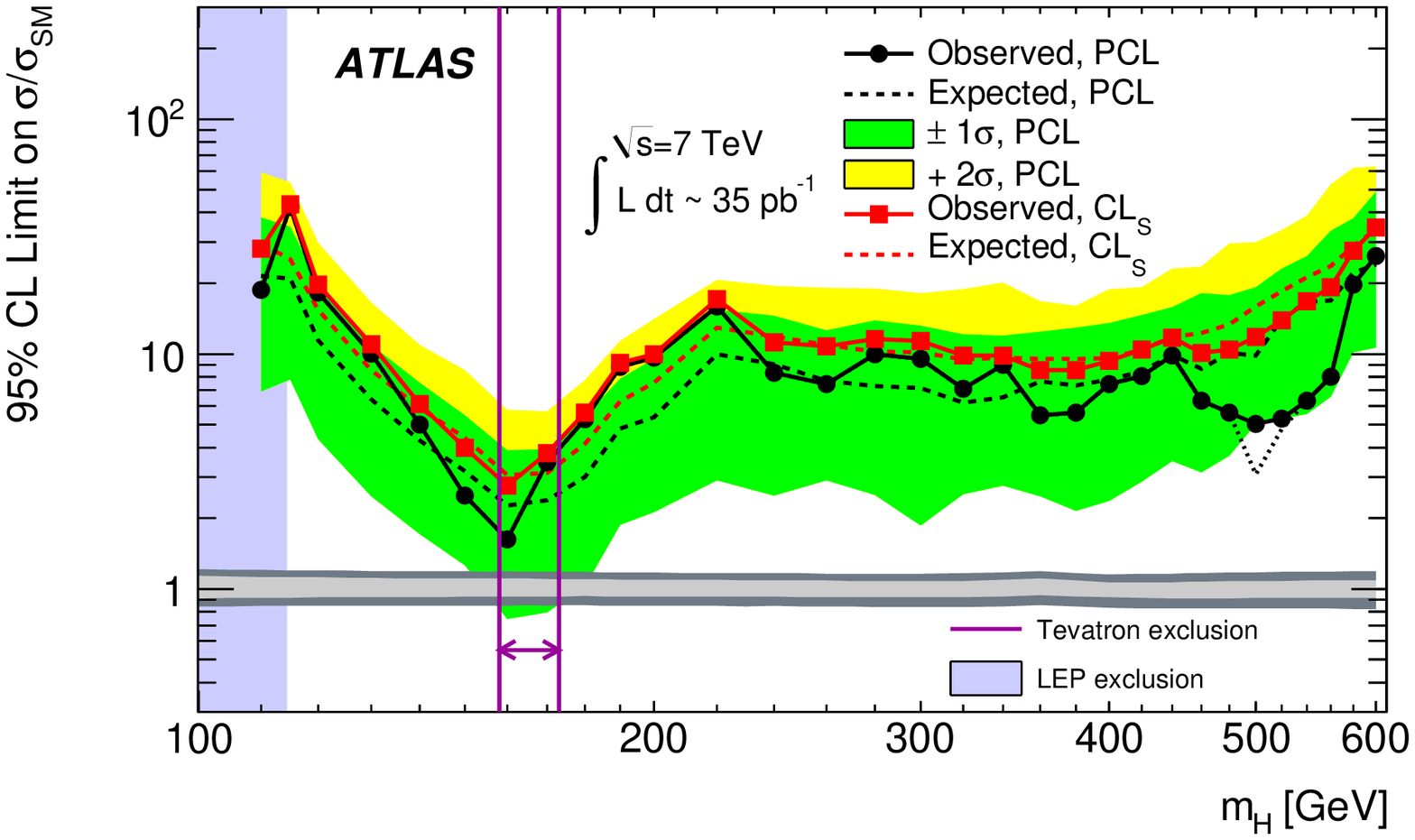}}}
\caption{Same as Fig.~\ref{fig:limit-div}, except that limits calculated using the CL$_{s}$ procedure are
  added. As expected, when the observed limits fluctuate up, both
  methods converge, but
  downward fluctuations are less pronounced with CL$_{s}$  due to its  larger
  over-coverage.
The fine dotted line marks the results obtained using CL$_{sb}$,
and the application of the power constraint gives the solid line.
The limits are calculated at the masses marked with symbols.
 The lines between the points are to guide the eye.
The regions excluded by the combined LEP
experiments~\cite{Barate:2003sz} and the 
Tevatron experiments~\cite{ref:tevhiggs} are indicated. 
\label{fig:limit-cls}
}
\end{figure*}

The excluded signal strength as a function of \mH\ is summarised in
Table~\ref{tab:all_excl}, using  the PCL method. 
The results are also calculated using the CL$_s$
method for comparison purposes:
the extracted limits from
both procedures are shown in Fig.~\ref{fig:limit-cls}.
Also given  is the profile likelihood ratio of a \SM\ Higgs boson to
background only and the consistency of the data with the 
background-only hypothesis,  $p_{\mu=0}$.

\begin{table*}[!htbp]
  \begin{center}
  \caption{The signal cross sections, in multiples of the
    \SM\ cross section, that are excluded, and expected to
    be excluded, at $95\%$~CL. 
The expected variation at $\pm1\sigma$ is also given for the PCL limits.
The bold numbers show the limit which
    should be used; for mass of  500 and 520~\GeV\ the power constraint is
    applied. The likelihood ratio of signal plus background to
    background is also shown, as is the p-value (modified to go between
    0 and 1) for $\mu=0$, which can
    be used to estimate the discovery significance.
\label{tab:all_excl}
}
\vspace*{0.4cm}
    \begin{tabular}{ccccccccccc}
      \hline
      \hline
\mH (\GeV) & \multicolumn{4}{c}{PCL limits}&
                         \multicolumn{2}{c}{CL$_s$ limits} & 
                 $-2\ln \frac{{\cal L}(1,\hat{\hat{\theta}})}
                        {{\cal L}(0,\hat{\hat{\theta}})}$
                                             & \multicolumn{2}{c}{p-values}  \\
          & Obs. & -1$\sigma$ & Median & +1$\sigma$ &   Obs. & Median &
                        & $p_{\mu=0}$  \\
      \hline

 110 & {\bf 18.7} &  6.9 & 21.5 & 38.2 & 28.1 & 29.6 & 0.1 &0.58 \\
 115 & {\bf 42.4} &  7.8 & 20.9 & 34.7 & 43.5 & 25.3 &-0.3 &0.07 \\
 120 & {\bf 18.2} &  4.3 & 11.4 & 19.9 & 19.7 & 15.4 &-0.3 &0.22 \\
 130 & {\bf 10.0} &  2.5 &  6.4 & 10.9 & 11.0 &  8.5 &-0.6 &0.23 \\
 140 & {\bf  5.0} &  1.7 &  4.3 &  7.6 &  6.1 &  5.9 & 0.0 &0.41 \\
 150 & {\bf  2.5} &  1.3 &  3.2 &  5.5 &  4.0 &  4.4 & 1.0 &0.65 \\
 160 & {\bf  1.6} &  0.7 &  2.3 &  3.9 &  2.8 &  3.1 & 1.6 &0.68 \\
 170 & {\bf  3.4} &  0.8 &  2.4 &  3.9 &  3.8 &  3.1 &-0.4 &0.27 \\
 180 & {\bf  5.3} &  1.0 &  3.0 &  5.4 &  5.6 &  4.2 &-0.9 &0.18 \\
 190 & {\bf  8.8} &  1.9 &  4.8 &  7.8 &  9.2 &  6.3 &-1.1 &0.11 \\
 200 & {\bf  9.7} &  2.1 &  5.4 &  9.5 &  9.9 &  7.5 &-1.1 &0.15 \\
 220 & {\bf 15.9} &  2.9 & 10.0 & 15.4 & 17.1 & 12.9 & 0.0 &0.13 \\
 240 & {\bf  8.3} &  2.5 &  9.1 & 14.5 & 11.2 & 11.9 & 0.3 &0.57 \\
 260 & {\bf  7.4} &  2.9 &  7.7 & 12.6 & 10.8 & 10.9 & 0.4 &0.56 \\
 280 & {\bf 10.0} &  2.5 &  7.3 & 13.9 & 11.5 & 10.2 & 0.2 &0.31 \\
 300 & {\bf  9.5} &  1.9 &  7.2 & 13.2 & 11.4 & 10.1 & 0.2 &0.32 \\
 320 & {\bf  7.1} &  2.5 &  6.2 & 12.1 &  9.8 &  9.5 & 0.4 &0.40 \\
 340 & {\bf  9.0} &  2.8 &  6.5 & 11.9 &  9.9 &  9.6 & 0.4 &0.28 \\
 360 & {\bf  5.5} &  2.5 &  7.7 & 12.5 &  8.5 &  9.5 & 0.4 &0.63 \\
 380 & {\bf  5.6} &  2.2 &  7.3 & 12.9 &  8.6 &  9.5 & 0.4 &0.53 \\
 400 & {\bf  7.5} &  2.4 &  7.8 & 13.6 &  9.4 &  9.6 & 0.1 &0.49 \\
 420 & {\bf  8.0} &  2.9 &  8.4 & 14.7 & 10.4 & 10.4 & 0.2 &0.46 \\
 440 & {\bf  9.8} &  3.5 &  9.9 & 15.8 & 11.7 & 11.8 & 0.1 &0.46 \\
 460 & {\bf  6.3} &  3.1 &  8.7 & 18.1 & 10.1 & 12.3 & 0.4 &0.64 \\
 480 & {\bf  5.6} &  3.7 & 10.1 & 17.7 & 10.4 & 13.4 & 0.4 &0.80 \\
 500 &  3.1 & {\bf  5.0} &  9.9 & 19.3 & 11.8 & 15.9 & 0.5 &0.89 \\
 520 &  4.8 & {\bf  5.3} & 14.1 & 23.1 & 13.9 & 18.5 & 0.4 &0.86 \\
 540 & {\bf  6.3} &  5.6 & 16.7 & 26.0 & 16.8 & 21.3 & 0.4 &0.82 \\
 560 & {\bf  8.0} &  6.6 & 16.9 & 33.5 & 19.3 & 23.7 & 0.3 &0.80 \\
 580 & {\bf 19.7} & 10.2 & 22.2 & 37.9 & 27.5 & 28.8 & 0.1 &0.56 \\
 600 & {\bf 26.1} & 10.6 & 24.0 & 49.2 & 34.4 & 33.9 & 0.1 &0.45 \\
      \hline \hline
    \end{tabular}
  \end{center}
\end{table*}

\begin{figure*}[!p]
\centering  
\mbox{\hspace*{0.0cm}{\includegraphics[width=0.85\textwidth]{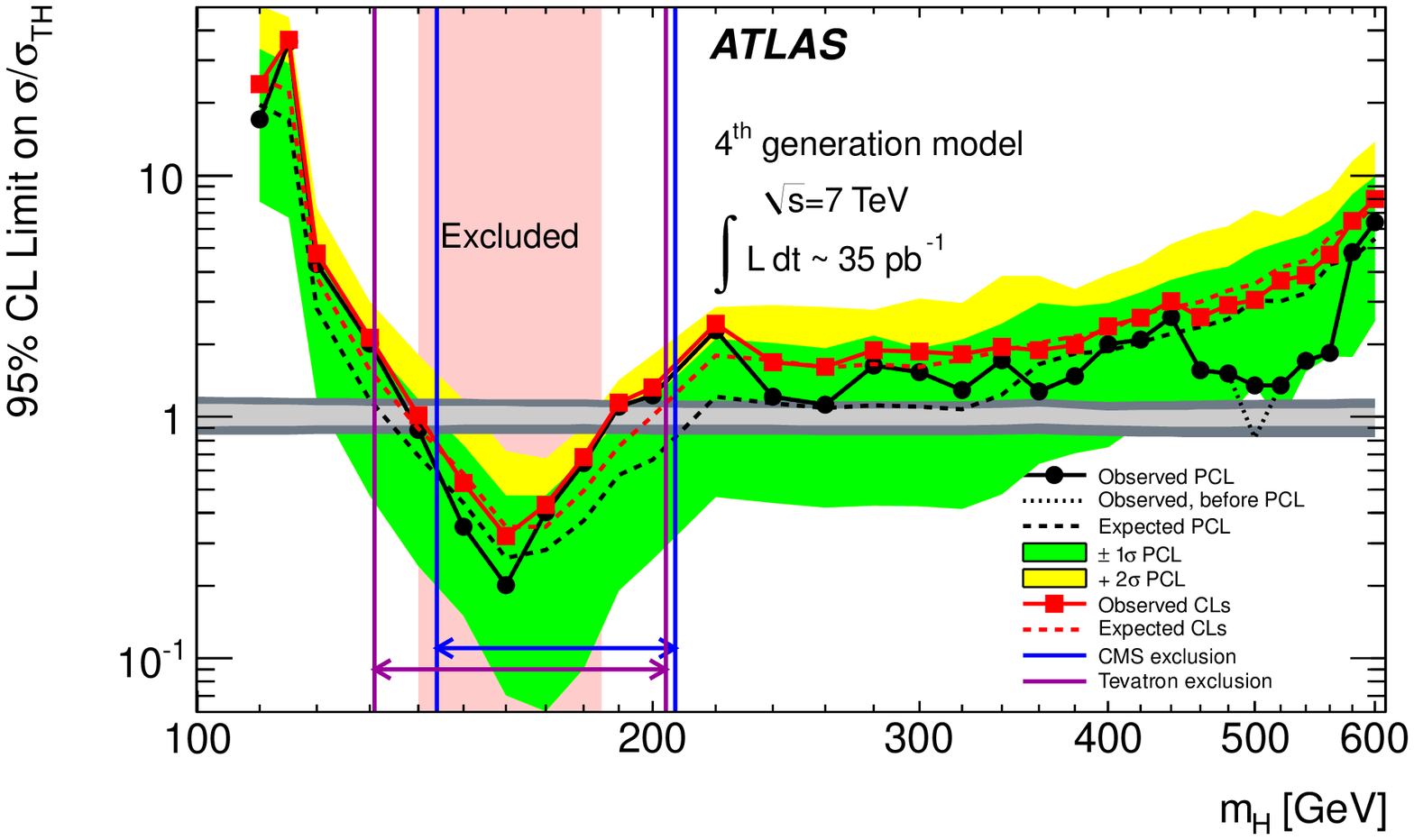}}}
\caption{Same as Fig.~\ref{fig:limit-div}, but comparing the excluded
cross section to the expected one when a fourth generation of high
mass quarks and leptons
with \SM-like couplings to the Higgs boson are included in the cross section calculations. 
The arrows indicate
  the regions excluded by the CMS experiment~\cite{Chatrchyan:2011tz}
  and the Tevatron 
  experiments~\cite{Aaltonen:2010sv}. 
The point set at 500~\GeV\ is at the limit allowed by  the power
constraint. 
The limits are calculated at the masses
 marked with symbols. 
 The lines between the points are to guide the eye.
\label{fig:limit-4g}
}
\end{figure*}

The results have  been interpreted in terms of the heavy mass fourth
generation model introduced in Section~\ref{sec:higgsxsbr}. This
involves rescaling the gluon fusion component of the production
cross section and  the Higgs boson decay branching ratios. The
limits are shown
in Fig.~\ref{fig:limit-4g}. This model is excluded for Higgs boson
masses between 140~GeV and 185~\GeV, while the region in which
exclusion might be expected is between 136~GeV and 208~\GeV.

The excluded cross section ratios are summarised in
Table~\ref{tab:all_excl4}.

\begin{table*}[!p]
  \begin{center}
  \caption{
The  signal cross sections, in multiples of the high mass fourth
generation model~\cite{ruan_zhang} cross section, that are excluded,
and expected to be excluded, in the absence of signal, at 95\% CL. 
\label{tab:all_excl4}
}
\vspace*{0.2cm}
    \begin{tabular}{cccccc|cccccc}
      \hline
      \hline
\mH (\GeV) & \multicolumn{3}{c}{PCL limits}&
                         \multicolumn{2}{c|}{CL$_s$ limits} &
\mH (\GeV) & \multicolumn{3}{c}{PCL limits}&
                         \multicolumn{2}{c}{CL$_s$ limits} \\
       &      Obs.    & -1$\sigma$ & Median  & Median & Obs.& 
       &      Obs.    & -1$\sigma$ & Median  & Median & Obs.\\
      \hline
 110 & {\bf 17.1} &        7.8 & 19.6 & 25.2 & 23.9 &
 320 &  {\bf 1.3} &        0.4 & 1.1 & 1.7 &  1.8 \ \\
 115 & {\bf 35.8} &        6.7 & 17.0 & 22.4 & 36.6 &
 340 &  {\bf 1.7} &        0.5 & 1.2 & 1.9 &  1.9 \\
 120 &  {\bf 4.3} &       1.2  & 2.8  &  3.8 &  4.8 &
 360 &  {\bf 1.3} &        0.6 & 1.7 & 2.0 &  1.9  \\
 130 &  {\bf 2.0} &       0.5  & 1.2  & 1.6 &  2.1 &
 380 &  {\bf 1.5} &        0.7 & 1.8 & 2.1 &  2.0  \\
 140 &  {\bf 0.9} &        0.2 & 0.7 &  0.9 &  1.0  &
 400 &  {\bf 2.0} &        0.8 & 1.9 & 2.2 &  2.3  \\
 150 &  {\bf 0.4} &        0.2 &  0.4 & 0.6 &  0.5 &
 420 &  {\bf 2.1} &        0.9 & 2.0 & 2.6 &  2.5  \\
 160 &  {\bf 0.2} &        0.1 & 0.3 & 0.4  & 0.3  &
 440 &  {\bf 2.6} &        1.1 & 2.2 & 2.8 &  3.0\\
 170 &  {\bf 0.4} &        0.1 & 0.3 & 0.4  & 0.4 &
 460 &  {\bf 1.6} &        1.2 & 2.4 & 3.0 &  2.6 \\
 180 &  {\bf 0.6} &        0.1 & 0.4 & 0.5 &  0.7 &
 480 &  {\bf 1.5} &        1.0 & 2.5 & 3.3 &  2.9 \\
 190 &  {\bf 1.1} &        0.2 & 0.6 & 0.8 & 1.1 &
 500 &  0.8       &  {\bf 1.4} & 3.0 & 3.5 &  3.0  \\
 200 &  {\bf 1.2} &        0.3 & 0.7 & 1.0 &  1.3  &
 520 &  {\bf 1.3} &        0.9 & 3.0 & 4.1 &  3.6 \\
 220 &  {\bf 2.3} &        0.5 & 1.2 & 1.8 &  2.4 &
 540 &  {\bf 1.7} &        1.5 & 3.3 &  4.3 &  3.8 \\
 240 &  {\bf 1.2} &        0.4 & 1.1 &  1.7 &  1.7 &
 560 &  {\bf 1.8} &        1.8 & 4.3 & 5.4 &  4.6  \\
 260 &  {\bf 1.1} &        0.4 & 1.1 &  1.6 &  1.6 &
 580 &  {\bf 4.8} &        1.8 & 4.6 & 6.0 &  6.3  \\
 280 &  {\bf 1.6} &        0.4 & 1.1 & 1.6 &  1.9 &
 600 &  {\bf 6.4} &        2.5 & 5.5 & 7.4 &  7.8  \\
 300 &  {\bf 1.5} &        0.4 & 1.1 & 1.6 &  1.9 \\

      \hline \hline
    \end{tabular}
  \end{center}
\end{table*}

\section{Conclusions}
\label{sec:conc}

The  ATLAS  search for the \SM\ Higgs boson
in the mass range from 110~\GeV\ to 600~\GeV\ in 35 to 40 \ipb\ of
    data recorded in  2010   has been presented. 
With this luminosity there is not sufficient sensitivity to
exclude the \SM\ Higgs boson, nor is there any evidence of an excess
of events over the predicted background rates. 
However, these results
give the most stringent constraints to date for Higgs boson masses above
250~\GeV\ and are close 
to the Tevatron limits~\cite{ref:tevhiggs} at intermediate masses.

An extension of the \SM, assuming the Higgs mechanism and adding a
 fourth generation of heavy quarks and 
leptons, is excluded for Higgs boson masses between 140 and 185~\GeV.

\section*{Acknowledgements \label{sec:Ack}}

We thank CERN for the very successful operation of the LHC, as well as the
support staff from our institutions without whom ATLAS could not be
operated efficiently.

We acknowledge the support of ANPCyT, Argentina; YerPhI, Armenia; ARC,
Australia; BMWF, Austria; ANAS, Azerbaijan; SSTC, Belarus; CNPq and FAPESP,
Brazil; NSERC, NRC and CFI, Canada; CERN; CONICYT, Chile; CAS, MOST and
NSFC, China; COLCIENCIAS, Colombia; MSMT CR, MPO CR and VSC CR, Czech
Republic; DNRF, DNSRC and Lundbeck Foundation, Denmark; ARTEMIS, European
Union; IN2P3-CNRS,\\
 CEA-DSM/IRFU, France; GNAS, Georgia; BMBF, DFG, HGF, MPG
and AvH Foundation, Germany; GSRT, Greece; ISF, MINERVA, GIF, DIP and
Benoziyo Center, Israel; INFN, Italy; MEXT and JSPS, Japan; CNRST, Morocco;
FOM and NWO, Netherlands; RCN, Norway; MNiSW, Poland; GRICES and FCT,
Portugal; MERYS (MECTS), Romania; MES of Russia and ROSATOM, Russian
Federation; JINR; MSTD, Serbia; MSSR, Slovakia; ARRS and MVZT, Slovenia;
DST/NRF, South Africa; MICINN, Spain; SRC and Wallenberg Foundation,
Sweden; SER, SNSF and Cantons of Bern and Geneva, Switzerland; NSC, Taiwan;
TAEK, Turkey; STFC, the Royal Society and Leverhulme Trust, United Kingdom;
DOE and NSF, United States of America.

The crucial computing support from all WLCG partners is acknowledged
gratefully, in particular from CERN and the ATLAS Tier-1 facilities at
TRIUMF (Canada), NDGF (Denmark, Norway, Sweden), CC-IN2P3 (France),
KIT/GridKA (Germany), INFN-CNAF (Italy), NL-T1\\ (Netherlands), PIC (Spain),
ASGC (Taiwan), RAL (UK) and BNL (USA) and in the Tier-2 facilities
worldwide.

\clearpage

\bibliographystyle{atlasnote}
\bibliography{paper}

\clearpage
\onecolumn
\begin{flushleft}
{\Large The ATLAS Collaboration}

\bigskip

G.~Aad$^{\rm 48}$,
B.~Abbott$^{\rm 111}$,
J.~Abdallah$^{\rm 11}$,
A.A.~Abdelalim$^{\rm 49}$,
A.~Abdesselam$^{\rm 118}$,
O.~Abdinov$^{\rm 10}$,
B.~Abi$^{\rm 112}$,
M.~Abolins$^{\rm 88}$,
H.~Abramowicz$^{\rm 153}$,
H.~Abreu$^{\rm 115}$
% ...
\bigskip

% ...
$^{10}$ Institute of Physics, Azerbaijan Academy of Sciences, Baku, Azerbaijan\\
$^{11}$ Institut de F\'isica d'Altes Energies and Universitat Aut\`onoma  de Barcelona and ICREA, Barcelona, Spain\\
% ...
$^{48}$ Fakult\"{a}t f\"{u}r Mathematik und Physik, Albert-Ludwigs-Universit\"{a}t, Freiburg i.Br., Germany\\
$^{49}$ Section de Physique, Universit\'e de Gen\`eve, Geneva, Switzerland\\
% ...
$^{88}$ Department of Physics and Astronomy, Michigan State University, East Lansing MI, United States of America\\
% ...
$^{111}$ Homer L. Dodge Department of Physics and Astronomy, University of Oklahoma, Norman OK, United States of America\\
$^{112}$ Department of Physics, Oklahoma State University, Stillwater OK, United States of America\\
% ...
$^{115}$ LAL, Univ. Paris-Sud and CNRS/IN2P3, Orsay, France\\
% ...
$^{118}$ Department of Physics, Oxford University, Oxford, United Kingdom\\
% ...
$^{153}$ Raymond and Beverly Sackler School of Physics and Astronomy, Tel Aviv University, Tel Aviv, Israel\\
% ...
\end{flushleft}


\end{document}